\documentclass[prc,letterpaper,twocolumn,showpacs,showkeys,lengthcheck,tightenlines,
               nofootinbib,preprintnumbers,superscriptaddress]{revtex4-1}

\usepackage[utf8]{inputenc}

\usepackage{titlesec}
\usepackage{dcolumn}
\usepackage{bm}
\usepackage{amsmath,amssymb,amsfonts}

\usepackage{graphicx}
\usepackage{subfigure}
\usepackage{color}
\usepackage{bbold}

\usepackage{slashed}
\usepackage[svgnames]{xcolor}
\usepackage[english]{babel}
\usepackage{blindtext}
\usepackage{microtype}
\usepackage{tikz}

\usepackage[colorlinks=true,urlcolor=blue,citecolor=blue]{hyperref}
\usepackage{ifpdf}
\usepackage{float}

\graphicspath{{.}{figures/}}

\titlespacing{\section}{5pt}{12pt plus 4pt minus 2pt}{8pt plus 2pt minus 2pt}
\titlespacing{\subsection}{0pt}{12pt plus 4pt minus 2pt}{8pt plus 2pt minus 2pt}

\begin{document}

\title{Off-shell modifications of the pion generalized parton distributions and transverse momentum dependent parton distributions}

\author{Jin-Li Zhang}
\email[]{jlzhang@njit.edu.cn}
\affiliation{School of Mathematics and Physics, Nanjing Institute of Technology, Nanjing 211167, China }

\begin{abstract}
The off-shell characteristics of pion generalized parton distributions (GPDs) and transverse momentum dependent parton distributions (TMDs) are examined within the framework of the Nambu–Jona-Lasinio model. In our previous papers, we separately investigated the properties of on-shell pion GPDs and light-front wave functions. It is particularly intriguing to compare the differences between on-shell and off-shell pion GPDs, which allows us to explore the effects associated with off-shellness. Due to the absence of crossing symmetry, the moments of GPDs also incorporate odd powers of the skewness parameter, resulting in new off-shell form factors. Through our calculations, we derived correction functions that account for modifications in pion GPDs due to off-shell effects. Unlike their on-shell counterparts, certain properties break down in the off-shell scenario; for instance, symmetry properties and polynomiality conditions may no longer hold. Additionally, we evaluate off-shell TMDs and compare them with their on-shell equivalents while also investigating their dependence on $\bm{k}_{\perp}$.
\end{abstract}


\maketitle
\section{Introduction}
For a thorough understanding of hadron structure, one may investigate generalized transverse momentum-dependent parton distributions (GTMDs)~\cite{Zhang:2020ecj,Puhan:2025kzz,Zhang:2024adr}, which are often referred to as the mother distributions. As the foundational distributions, GTMDs can be reduced to GPDs~\cite{Mueller:1998fv,Ji:1996nm,Radyushkin:1997ki,Ji:1998pc,Diehl:2003ny,Zhang:2021mtn,Zhang:2021shm,Zhang:2021tnr,Zhang:2021uak,Zhang:2022zim,Deja:2023tuc,Sun:2018ldr,Fu:2022bpf,Adhikari:2021jrh} and TMDs~\cite{Liu:2025fuf,Zhang:2024plq} under specific limits. GPDs and TMDs offer a comprehensive array of information regarding the confined spatial distributions of quarks and gluons within bound hadrons. GPDs, which represent the three-dimensional extension of conventional parton distribution functions (PDFs)~\cite{Tanisha:2025qda,Cheng:2024gyv,Wu:2025rto}, are defined as the off-forward matrix elements associated with quark and gluon operators. GPDs are observables that encompass a wealth of previously inaccessible information regarding hadron structure. They arise in exclusive processes such as deeply virtual Compton scattering (DVCS)~\cite{Castro:2025rpx} and timelike Compton scattering (TCS). Recently, a novel process known as double deeply virtual Compton scattering (DDVCS) has been introduced~\cite{Deja:2023ahc,Deja:2023tuc}. In contrast to DVCS, the DDVCS process involves an electron scattering off a nucleon, resulting in the production of a lepton pair. A notable feature of DDVCS is its potential for direct measurement of GPDs at $x \neq \pm \xi$ at leading order. When QCD factorization is applicable, the amplitude of high-energy processes can be expressed as a convolution of a hard perturbative kernel and GPDs. GPDs provide more comprehensive information than PDFs and form factors (FFs)~\cite{Zhang:2024nxl,Hernandez-Pinto:2024kwg,Bondarenko:2025qch} regarding the internal structure of hadrons. For instance, they encapsulate insights into the spin contributions from quarks and gluons within nucleons, as demonstrated in Refs.~\cite{Ji:1996ek,Ji:1996nm}. Consequently, GPDs serve as valuable tools for elucidating the transverse spatial distribution of partons. 

Given their significance in understanding hadronic structure dynamics, extensive research on GPDs has been conducted; comprehensive results can be found in reviews~\cite{Diehl:2003ny,Belitsky:2005qn,Mueller:2014hsa}. However, measuring GPDs experimentally remains challenging; consequently, first-principles computations using lattice quantum chromodynamics (QCD) have primarily focused on their lowest Mellin moments. Therefore, enhancing our theoretical understanding of GPD behavior would facilitate more accurate experimental determinations.

In the Sullivan process~\cite{Sullivan:1971kd}, the off-shell characteristics of the GPDs of pions are taken into account. The presence of off-shellness disrupts crossing symmetry, leading to the emergence of new off-shell FFs. In this study, we will investigate the off-shell GPDs and examine the relationship between these off-shell gravitational form factors of pions.

The investigation into the transverse momentum distribution of hadrons produced in semi-inclusive deep inelastic scattering (SIDIS)~\cite{Boer:1997nt,Collins:2007ph,Amrath:2005gv} is characterized by determining TMDs. TMDs have been extensively studied~\cite{Noguera:2015iia,Lorce:2014hxa,Matevosyan:2011vj,Puhan:2023ekt,Boglione:2024dal}. In this paper, we will also evaluate off-shell pion TMDs.

Pions are Nambu-Goldstone bosons associated with the chiral symmetry breaking in QCD among all hadrons, and they are believed to play a crucial role in the origin of mass and matter~\cite{Aguilar:2019teb,Roberts:2021nhw}. Therefore, understanding how quarks and gluons combine to form pions is of paramount importance; thus, gaining experimental insights into their structure would be highly valuable. Given its connection to chiral symmetry breaking, investigating the GPDs of pions is particularly significant.

In this study, we will calculate the off-shell GPDs of pions within the framework of the Nambu-Jona-Lasinio (NJL) model~\cite{RevModPhys.64.649,Buballa:2003qv,Zhang:2018ouu,Zhang:2016zto,Cui:2017ilj,Cui:2016zqp,Li:2018ltg,Zhang:2024dhs}. The NJL model is a well-established phenomenological approach to quark matter that incorporates essential QCD features such as chiral phase transitions along with various interaction terms that describe both quark dynamics and their interactions. It can accurately predict meson masses and decay constants while also playing a vital role in characterizing other properties of quark matter. Previous studies have already examined on-shell GPDs~\cite{Chen:2006gg,Broniowski:2007si,Mezrag:2014jka,Mezrag:2016hnp,Zhang:2021shm,Chavez:2021llq,Raya:2021zrz,Zhang:2021uak,Zhang:2021mtn,Mezrag:2022pqk}, which will allow us to compare our findings on off-shell GPDs effectively.

This paper is structured as follows: In Sec.~\ref{nice}, we begin with a concise introduction to the NJL model. Subsequently, we outline the process for defining and calculating pion off-shell GPDs. In Sec.~\ref{well}, we examine and discuss the fundamental properties of off-shell GPDs, with particular emphasis on the FFs associated with these distributions. In Sec.~\ref{tmdoffshell}, we explore the off-shell TMDs of pions. Finally, a brief summary and discussion are presented in Sec.~\ref{excellent}.

\section{Off-shell GPDs}\label{nice}
The off-shell GPDs of pions are examined within the framework of the Spectral Quark Model (SQM) in the chiral limit, as discussed in Refs.~\cite{Broniowski:2023his,Broniowski:2022iip}. In this section, we calculate off-shell GPDs using the NJL model, and we compare our findings with the on-shell pion GPDs presented in Refs.~\cite{Zhang:2021shm,Zhang:2021uak}. 

\subsection{Nambu--Jona-Lasinio Model}\label{good}

The SU(2) flavor NJL Lagrangian,
\begin{align}\label{1}
\mathcal{L}&=\bar{\psi }\left(i\gamma ^{\mu }\partial _{\mu }-\hat{m}\right)\psi+ G_{\pi }[\left(\bar{\psi }\psi\right)^2-\left( \bar{\psi }\gamma _5 \vec{\tau }\psi \right)^2].
\end{align}
The expression $\hat{m}\equiv\text{diag}\left[m_u,m_d\right]$ denotes the current quark mass matrix. Under the assumption of isospin symmetry, we have $m_u = m_d = m$. The symbols $\vec{\tau}$ represent the Pauli matrices associated with isospin, while $G_{\pi}$ refers to the four-fermion coupling constant.

The dressed quark propagator in the NJL model is derived by solving the gap equation,
\begin{align}\label{bc2}
iS^{-1}(k)=iS_0^{-1}(k)-\sum_{\Omega}K_{\Omega}\Omega \int \frac{d^4l}{(2\pi)^4}\text{tr}[\bar{\Omega}i S(l)],
\end{align}
where $S_0^{-1}(k)={\not\!k}-m+i \varepsilon$ represents the bare quark propagator, and the trace is taken over Dirac, color, and isospin indices. The solution of gap equation is defined as 
\begin{align}\label{2}
S(k)=\frac{1}{{\not\!k}-M+i \varepsilon}.
\end{align}
The interaction kernel of the gap equation is local; therefore, we derive a constant dressed quark mass
\begin{align}\label{3}
M=m+12 i G_{\pi}\int \frac{d^4l}{(2 \pi )^4}\text{tr}_{\text{D}}[S(l)],
\end{align}
where the trace is taken over Dirac indices. Dynamical chiral symmetry breaking can occur only when the coupling strength exceeds a critical threshold, specifically $G_{\pi} > G_{critical}$, leading to a nontrivial solution where $M > 0$.

A regularization procedure is essential for the complete specification of the NJL model, as it constitutes a non-renormalizable quantum field theory. In Ref.~\cite{Zhang:2021uak}, we have examined the dependence of pion on-shell GPDs on the chosen regularization scheme within the context of the NJL model. In this paper, we adopt the proper time regularization (PTR) scheme~\cite{Hellstern:1997nv,Bentz:2001vc}.
\begin{align}\label{ptr}
\frac{1}{X^n}&=\frac{1}{(n-1)!}\int_0^{\infty}d\tau \tau^{n-1}e^{-\tau X}\nonumber\\
& \rightarrow \frac{1}{(n-1)!} \int_{1/\Lambda_{\text{UV}}^2}^{1/\Lambda_{\text{IR}}^2}d\tau \tau^{n-1}e^{-\tau X},
\end{align}
where $X$ represents a product of propagators that have been combined using Feynman parametrization. Beyond the ultraviolet cutoff, $\Lambda_{\text{UV}}$, we also introduce the infrared cutoff $\Lambda_{\text{IR}}$ to mimic confinement, it should be of the order $\Lambda_{\text{QCD}}$ and we choose $\Lambda_{\text{IR}}=0.240$ GeV. The parameters used in this work are given in Table \ref{tb1}. 

We will use the notations in Eqs. (\ref{cfun}) and (\ref{cfun1}) in the following.
\begin{center}
\begin{table}
\caption{Parameter set used in our work. The dressed quark mass and regularization parameters are in units of GeV, while coupling constant are in units of GeV$^{-2}$.}\label{tb1}
\begin{tabular}{p{0.9cm} p{0.9cm} p{0.8cm} p{0.9cm}p{0.9cm}p{0.9cm}p{0.9cm}p{0.9cm}}
\hline\hline
$\Lambda_{\text{IR}}$&$\Lambda _{\text{UV}}$&$M$&$G_{\pi}$&$Z_{\pi}$&$m_{\pi}$&$G_{\omega}$&$G_{\rho}$\\
\hline
0.240&0.645&0.4&19.0&17.85&0.14&10.4&11.0\\
\hline\hline
\end{tabular}
\end{table}
\end{center}

\subsection{The Definition and Calculation of the Pion Off-Shell GPDs}\label{qq}
\begin{figure}
\centering
\includegraphics[width=0.47\textwidth]{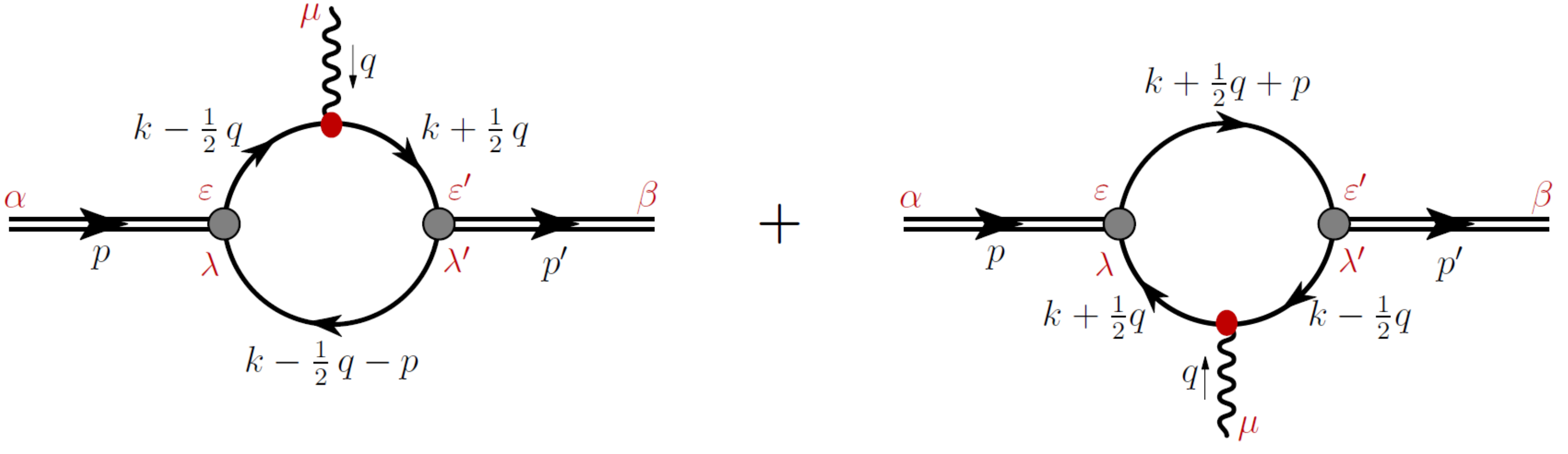}
\caption{Pion off-shell GPDs diagrams, different from the pion on-shell GPDs in Ref.~\cite{Zhang:2021shm}, here $p^2\neq p^{\prime2}\neq m_{\pi}^2$.}\label{GPD}
\end{figure}

The off-shell GPDs of the pion in the NJL model are illustrated in Fig. \ref{GPD}. In this context, $p$ represents the incoming pion momentum, while $p^{\prime}$ denotes the outgoing pion momentum. Unlike the on-shell case, we have $p^2 \neq p^{\prime2} \neq m_{\pi}^2$. In this paper, we will adopt the symmetry notation as used in Refs.~\cite{Ji:1998pc,Diehl:2003ny}. The kinematics of this process and related quantities are defined as follows:
\begin{align}\label{4}
t= q^2=(p'-p)^2=-Q^2,
\end{align}
\begin{align}\label{5}
 \xi=\frac{p^+-p^{\prime+}}{p^++p^{\prime+}}, \quad P=\frac{p+p^{\prime}}{2}, 
\end{align}
where $\xi$ represents the skewness parameter, expressed in light-cone coordinates
\begin{align}\label{4A}
v^{\pm}&=(v^0\pm v^3), \quad  \mathbf{v}=(v^1,v^2),
\end{align}
for any four-vector, $n=(1,0,0,-1)$ is the light-cone four-vector, then $v^+$ in the light-cone coordinate can be expressed as follows:
\begin{align}\label{4B}
v^+=v\cdot n.
\end{align}
These coordinates provide a natural framework for describing the infinite momentum frame, within which parton distributions can be elucidated through the physical perspective of the parton model.

The two leading-twist quark off-shell GPDs of the pion are defined as follows:
\begin{align}\label{dgpd}
&H(x,\xi,t,p^2,p^{\prime2})=\frac{1}{2}\int \frac{dz^-}{2\pi}e^{\frac{i}{2}x(p^++p^{\prime+})z^-}\nonumber\\
&\times \langle p^{\prime}|\bar{q}(-\frac{1}{2}z)\gamma^+q(\frac{1}{2}z)|p\rangle \mid_{z^+=0,\bm{z}=\bm{0}},
\end{align}
\begin{align}\label{dtgpd}
&\frac{P^+q^j-P^jq^+}{P^+m_{\pi}}E(x,\xi,t,p^2,p^{\prime2})=\frac{1}{2}\int \frac{dz^-}{2\pi}e^{\frac{i}{2}x(p^++p^{\prime+})z^-}\nonumber\\
&\times \langle p^{\prime}|\bar{q}(-\frac{1}{2}z)i\sigma^{+j}q(\frac{1}{2}z)|p\rangle \mid_{z^+=0,\bm{z}=\bm{0}},
\end{align}
where $x$ is the longitudinal momentum fraction. The first is the vector (no spin flip) off-shell GPD, the second is tensor (spin flip) off-shell GPD.
The operators for the two off-shell GPDs in Fig. \ref{GPD} read
\begin{subequations}\label{a91}
\begin{align}\label{6A}
\textcolor{red}{\bullet}_1 &=\gamma^+\delta(x-\frac{k^++k^{\prime+}}{p^++p^{\prime+}})\,, \\
\textcolor{red}{\bullet}_2 &=i\sigma^{+j} \delta(x-\frac{k^++k^{\prime+}}{p^++p^{\prime+}}),
\end{align}
\end{subequations}
$\textcolor{red}{\bullet}_1$ for vector off-shell GPD and $\textcolor{red}{\bullet}_2$ for tensor off-shell GPD. The pion vertex function, in the light-cone normalization, is given by
\begin{align}\label{6C}
\Gamma_{\pi}=\sqrt{Z_{\pi}}\gamma_5,
\end{align}
where $Z_{\pi}$ is the square of the effective meson-quark-quark coupling constant.

In the NJL model the off-shell GPDs can be written as
\begin{align}\label{gpddd}
&H\left(x,\xi,t,p^2,p^{\prime 2}\right)
=2i N_c Z_{\pi} \int \frac{d^4k}{(2 \pi )^4}\delta_n^x (k)\nonumber\\
&\times \text{tr}_{\text{D}}\left[\gamma _5 S \left(k_{+q}\right)\gamma ^+ S\left(k_{-q}\right)\gamma _5 S\left(k-P\right)\right],
\end{align}
\begin{align}\label{tgpddd}
& \frac{P^+q^j-P^jq^+}{P^+m_{\pi}} E\left(x,\xi,t,p^2,p^{\prime 2}\right)=2i N_c Z_{\pi}\int \frac{d^4k}{(2 \pi )^4}\delta_n^x (k)\nonumber\\
&\times \text{tr}_{\text{D}}\left[\gamma _5 S \left(k_{+q}\right)i\sigma^{+j}S\left(k_{-q}\right)\gamma _5 S\left(k-P\right)\right],
\end{align}
where $\text{tr}_{\text{D}}$ indicates a trace over spinor indices, $\delta_n^x (k)=\delta (xP^+-k^+)$, $k_{+q}=k+\frac{q}{2}$, $k_{-q}=k-\frac{q}{2}$.
Here we use the following reduce formulae ($D(k^2)=k^2-M^2$)
\begin{subequations}\label{dr}
\begin{align}
p\cdot q&=\frac{p^{\prime2}-p^2-q^2}{2}\,, \\
k\cdot q&=\frac{1}{2} \left(D(k_{+q}^2)-D(k_{-q}^2)\right)\,, \\
k\cdot p&=-\frac{1}{2} \left(D((k-P)^2)-D(k_{-q}^2)-\frac{p^{\prime2}+p^2-q^2}{2}\right)\,, \\
k^2&=\frac{1}{2} \left(D(k_{+q}^2)+D(k_{-q}^2)\right)+M^2-\frac{q^2}{4},
\end{align}
\end{subequations}
to incorporate these relationships into Eqs. (\ref{gpddd}) and (\ref{tgpddd}), we first cancel each identical factor in the numerators and denominators. Subsequently, by applying Feynman parametrizations to simplify all remaining denominators, we arrive at the following result:
\begin{align}\label{agpdf}
&\quad H(x,\xi,t,p^2,p^{\prime 2})\nonumber\\
&=\frac{N_cZ_{\pi }}{8\pi ^2} \left[\theta_{\bar{\xi} 1}\bar{\mathcal{C}}_1(\sigma_4)+ \theta_{\xi 1} \bar{\mathcal{C}}_1(\sigma_5)+\theta_{\bar{\xi} \xi}\frac{x}{\xi }\bar{\mathcal{C}}_1(\sigma_6)\right]\nonumber\\
&+\frac{N_c Z_{\pi } }{8\pi ^2}\int_0^1 d\alpha \frac{\bar{\mathcal{C}}_2(\sigma_7)}{\sigma_7}\nonumber\\
&\times \frac{\theta_{\alpha \xi}\left((p^{\prime 2}-p^2)\xi+x(p^2+p^{\prime 2})+(1- x)t\right)}{\xi}  ,
\end{align}
\begin{align}\label{agtpdf}
E(x,\xi,t,p^2,p^{\prime 2})=\frac{N_c Z_{\pi } }{4\pi ^2}\int_0^1 d\alpha \frac{\theta_{\alpha \xi}}{\xi}m_{\pi}M\frac{\bar{\mathcal{C}}_2(\sigma_7)}{\sigma_7},
\end{align}
and
\begin{subequations}\label{region1}
\begin{align}
\theta_{\bar{\xi} 1}&=x\in[-\xi, 1]\,, \\
\theta_{\xi 1}&=x\in[\xi, 1]\,, \\
\theta_{\bar{\xi} \xi}&=x\in[-\xi, \xi]\,, \\
\theta_{\alpha \xi}&=x\in[\alpha (\xi +1)-\xi , \alpha  (1-\xi)+\xi ]\cap x\in[-1,1],
\end{align}
\end{subequations}
the step function $\theta$ indicates that $x$ exists solely within the corresponding region. One can express $\theta_{\bar{\xi} \xi}/\xi = \Theta(1 - x^2/\xi^2)$, where $\Theta(x)$ denotes the Heaviside function. Additionally, we have $\theta_{\alpha \xi}/\xi = \Theta((1 - \alpha^2) - (x - \alpha)^2/\xi^2)\Theta(1 - x^2)$. These results are valid in the region where $\xi > 0$. Under the transformation $\xi \rightarrow -\xi$, it follows that: $\theta_{\bar{\xi} 1} \leftrightarrow \theta_{\xi 1}$; and both $\theta_{\bar{\xi} \xi}/\xi$ and $\theta_{\alpha \xi}/\xi$ remain invariant.

\begin{figure}[t]
\includegraphics[clip,width=0.94\linewidth]{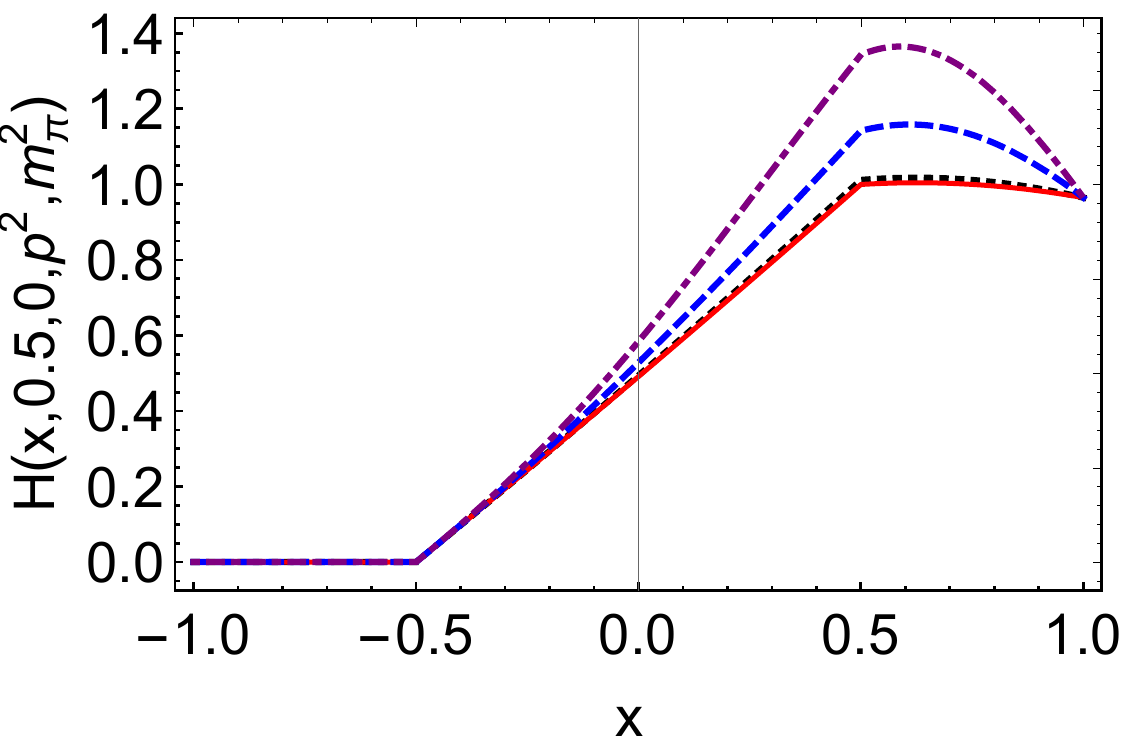}
\includegraphics[clip,width=0.94\linewidth]{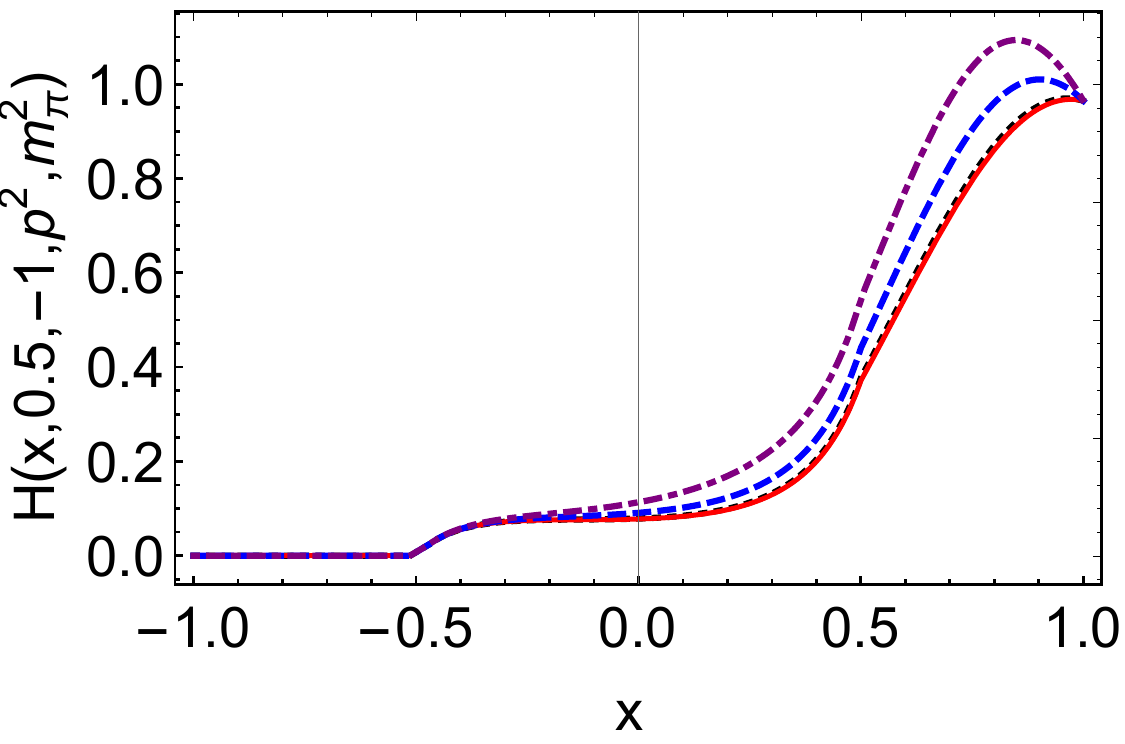}
\caption{\label{gpddgpd}
Pion off-shell vector GPD: $H(x,\xi,t,p^2,p^{\prime 2})$ in Eq. (\ref{agpdf}), we only plot $\xi>0$.
\emph{upper panel} -- The off-shell GPDs $H(x,0.5,0,p^2,m_{\pi}^2)$. $H(x,0.5,0,m_{\pi}^2,m_{\pi}^2)$ -- black dotted line; $H(x,0.5,0,0,m_{\pi}^2)$ -- red solid line; $H(x,0.5,0,0.2,m_{\pi}^2)$ -- blue dashed line; $H(x,0.5,0,0.4,m_{\pi}^2)$ -- purple dotdashed line.
\emph{lower panel} -- The off-shell GPDs $H(x,0.5,0,p^2,m_{\pi}^2)$. $H(x,0.5,-1,m_{\pi}^2,m_{\pi}^2)$ -- black dotted line; $H(x,0.5,-1,0,m_{\pi}^2)$ -- red solid line; $H(x,0.5,0,-1,0.2,m_{\pi}^2)$ -- blue dashed line; $H(x,0.5,-1,0.4,m_{\pi}^2)$ -- purple dotdashed line.
}
\end{figure}

\begin{figure}[t]
\includegraphics[clip,width=0.94\linewidth]{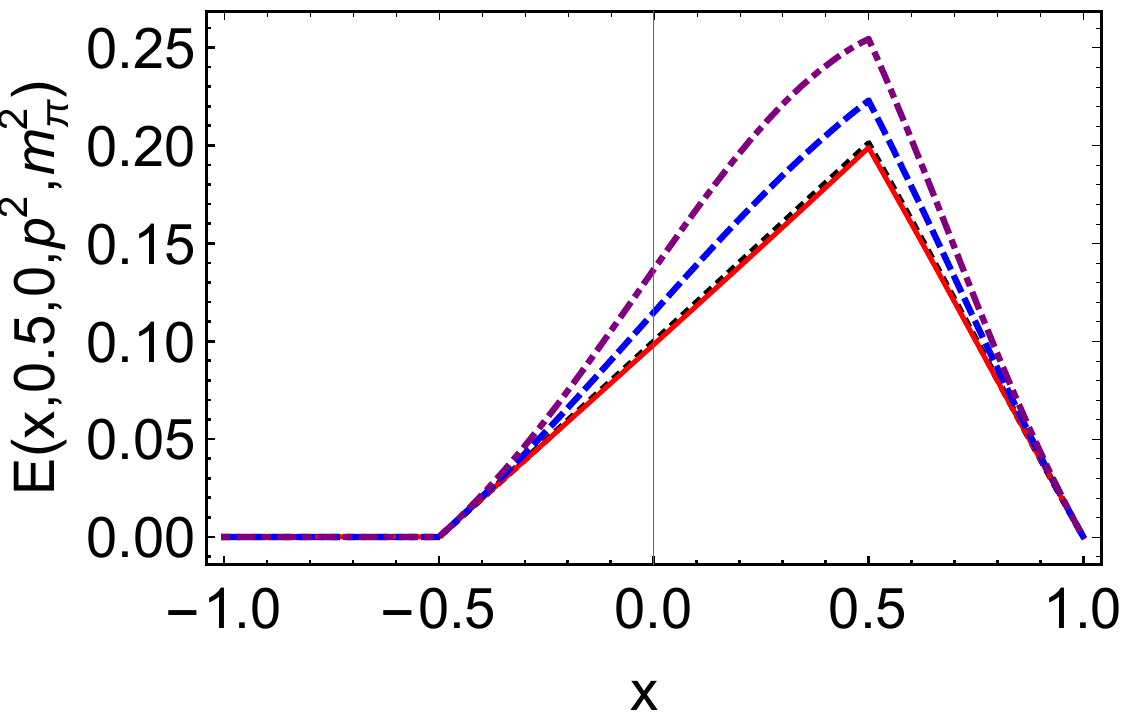}
\includegraphics[clip,width=0.94\linewidth]{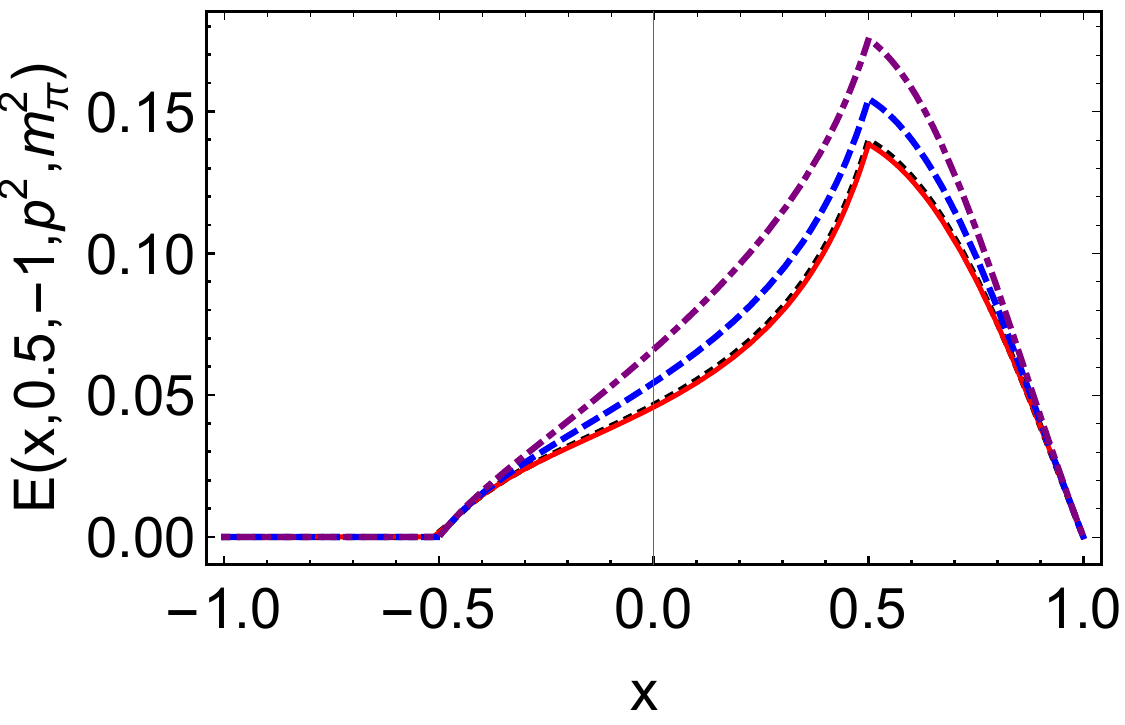}
\caption{\label{tgpddgpd}
Pion off-shell tensor GPD: $E(x,\xi,t,p^2,p^{\prime2})$ in Eq. (\ref{agtpdf}), we only plot $\xi>0$.
\emph{upper panel} -- The off-shell GPDs $E(x,0.5,0,p^2,m_{\pi}^2)$. $H(x,0.5,0,m_{\pi}^2,m_{\pi}^2)$ -- black dotted line; $E(x,0.5,0,0,m_{\pi}^2)$ -- red solid line; $E(x,0.5,0,0.2,m_{\pi}^2)$ -- blue dashed line; $E(x,0.5,0,0.4,m_{\pi}^2)$ -- purple dotdashed line.
\emph{lower panel} -- The off-shell GPDs $E(x,0.5,0,p^2,m_{\pi}^2)$. $E(x,0.5,-1,m_{\pi}^2,m_{\pi}^2)$ -- black dotted line; $E(x,0.5,-1,0,m_{\pi}^2)$ -- red solid line; $E(x,0.5,0,-1,0.2,m_{\pi}^2)$ -- blue dashed line; $E(x,0.5,-1,0.4,m_{\pi}^2)$ -- purple dotdashed line.
}
\end{figure}

Here, we present the diagrams of \( H(x,\xi,t,p^2,m_{\pi}^2) \) and \( E(x,\xi,t,p^2,m_{\pi}^2) \) in Figs. \ref{gpddgpd} and \ref{tgpddgpd}. The functions \( H(x,0.5,0,m_{\pi}^2,m_{\pi}^2) \) and \( E(x,0.5,0,m_{\pi}^2,m_{\pi}^2) \), which correspond to the on-shell GPDs, are discussed in Ref.~\cite{Zhang:2021shm}. In this work, we also plot the on-shell vector and tensor GPDs for comparative analysis.

From the diagrams, it is evident that when \( p^{\prime 2}=m_{\pi}^{2} \), the off-shellness depends on \( p^{2} \). As \( p^{2} \) increases, the off-shell effects of half-off-shell pion GPDs become more pronounced. Specifically, at \( p^{2}=0.2 \text{GeV}^2\), the relative effect is approximately $15 \%$ for the maximum value, while at a higher value of \( p^{2}=0.4 \text{GeV}^2\), this relative effect rises to about $25 \%$.



\section{The Properties of the Pion Off-shell GPDs}\label{well}

\subsubsection{FFs}
Unlike the on-shell case, time-reversal symmetry (or crossing symmetry) renders the GPDs an even function of $\xi$. However, for off-shell GPDs, this crossing symmetry no longer holds.

In general, the $x$-moments of the GPDs also incorporate odd powers of the skewness parameter $\xi$.
\begin{align}\label{pc}
\int_{-1}^1 x^n H(x,\xi,t,p^2,p^{\prime 2}) dx  &= \sum _{i=0}^{(n+1)} A_{n,i}(t,p,p^{\prime})\xi^i\,,
\end{align}
where $A_{n,i} $ are the generalized off-shell FFs. The most significant factors are the FFs associated with the electromagnetic and energy-stress tensor currents, as they do not depend on the factorization scale. The lowest $x$-moments
\begin{align}
\int_{-1}^1 x^0 H dx  = A_{1,0}+\xi A_{1,1}=F-G\xi,
\end{align}
\begin{align}
\int_{-1}^1 x^0 E dx  = B_{1,0}+\xi B_{1,1},
\end{align}
for $x^1$
\begin{align}\label{offgpd1}
\int_{-1}^1 x^1 H dx  &= A_{2,0}+\xi A_{2,1}+\xi^2 A_{2,2}\nonumber\\
&=\theta_2-\theta_3\xi-\theta_1\xi^2,
\end{align}
\begin{align}
\int_{-1}^1 x^1 E dx  = B_{2,0}+\xi B_{2,1}+\xi^2 B_{2,2},
\end{align}
where the FFs are functions of $(t,p^2,p^{\prime2})$. Unlike the on-shell FFs, the off-shell FFs exhibit dependence on both $t$ and the parameters associated with off-shellness. In comparison to the on-shell FFs discussed in Ref.~\cite{Zhang:2021shm}, it is noteworthy that $A_{1,1}$, $A_{2,1}$, $B_{1,1}$, and $B_{2,1}$ do not vanish. The general covariant structure of the pion-photon vertex is as follows:
\begin{align}
\Gamma^{\mu}(p,p^{\prime})=2P^{\mu}F(t,p^2,p^{\prime2})+q^{\mu}G(t,p^2,p^{\prime2})
\end{align}
at $t=0$, one can obtain the relationship~\cite{Broniowski:2022iip}
\begin{align}\label{g}
G(t,p^2,p^{\prime2})=\frac{(p^{\prime2}-p^2)}{t}\left[F(0,p^2,p^{\prime2})-F(t,p^2,p^{\prime2})\right],
\end{align}
and $G(0,p^2,p^{\prime2})=(p^{\prime2}-p^2)dF(t,p^2,p^{\prime2})/dt|_{t=0}$, one can obtain $G(t,p^2,p^2)=0$, which is a manifestation of the crossing symmetry. $F(t,p^2,p^{\prime2})$ is the electromagnetic FFs, and $F(0,m_{\pi}^2,m_{\pi}^2)=1$. From our off-shell GPDs, we obtain
\begin{align}\label{aF3}
A_{1,0}&=\frac{N_cZ_{\pi } }{8\pi^2}\int_0^1 dx\bar{\mathcal{C}}_1(\sigma_1)+\frac{N_cZ_{\pi } }{8\pi^2}\int_0^1 dx\bar{\mathcal{C}}_1(\sigma_2)\nonumber\\
&+\frac{N_cZ_{\pi }}{4\pi^2} \int_0^1 dx \int_0^{1-x}dy  \frac{1}{\sigma_8}\bar{\mathcal{C}}_2(\sigma_8) \nonumber\\
&\times ((p^2+p^{\prime 2})-(x+y)(p^{\prime 2}+p^2-t)) ,
\end{align}
\begin{align}\label{g1}
A_{1,1}&=\frac{N_cZ_{\pi } }{8\pi^2}\int_0^1 dx\bar{\mathcal{C}}_1(\sigma_1)-\frac{N_cZ_{\pi } }{8\pi^2}\int_0^1 dx\bar{\mathcal{C}}_1(\sigma_2)\nonumber\\
&-\frac{N_cZ_{\pi }}{4\pi^2} \int_0^1 dx \int_0^{1-x}dy  \frac{1}{\sigma_8}\bar{\mathcal{C}}_2(\sigma_8) (p^2-p^{\prime2}).
\end{align}
From the two equations presented above, we can deduce that \( F(t,p^2,p^{\prime 2}) = F(t,p^{\prime 2},p^2) \) and \( G(t,p^2,p^{\prime 2}) = -G(t,p^{\prime 2},p^2) \). This indicates that \( F(t,p^2,p^{\prime 2}) \) exhibits symmetry, while \( G(t,p^2,p^{\prime 2}) \) demonstrates antisymmetry.  

The study presented in Ref.~\cite{Choi:2019nvk} investigates the function $G(t,p^2,m_{\pi}^2)$ within the framework of a quark model. Our results are consistent with the result reported in Ref.~\cite{Choi:2019nvk}. Namely,  $-Q^2*G(t,p^{\prime2},p^2)$ and $(p^{\prime2}-p^2)(F(0,p^2,p^{\prime2})-F(t,p^2,p^{\prime2}))$ are in agreement with each other, which is consistent with our Eq. (\ref{g}) and Eq. (12) of Ref.~\cite{Broniowski:2022iip}.


We focus exclusively on the real parts of functions \( G \) and \( F \), as the PTR outlined in Eq. (\ref{ptr}) is only applicable for \( X > 0 \). When \( X \) includes an imaginary component, the formulation becomes significantly more complex. This issue has been addressed by Ref.~\cite{Kohyama:2015hix}, which provides a formula for cases where \( X \) contains an imaginary part. We intend to further explore off-shell FFs of pions in our subsequent work.


%

The parameters $\theta_1$, $\theta_2$, and $\theta_3$ are associated with the off-shell gravitational FFs. The general tensor structure of the gravitational vertex is expressed as follows:
\begin{align}
\Gamma^{\mu\nu} &= \frac{1}{2}[(q^2g^{\mu\nu}-q^{\mu}q^{\nu})\theta_1+4P^{\mu}P^{\nu}\theta_2\nonumber\\
&+2(q^{\mu}P^{\nu}+q^{\nu}P^{\mu})\theta_3-g^{\mu\nu}\theta_4],
\end{align}
under the cross symmetry, $\theta_1$ and $\theta_2$ are classified as even functions, while $\theta_3$ and $\theta_4$ are classified as odd functions. 

From Ref.~\cite{Broniowski:2022iip}, we know that
\begin{align}\label{theta3}
\theta_3(t,p^2,p^{\prime2})=\frac{(p^{\prime2}-p^2)}{t}\left[\theta_2(0,p^2,p^{\prime2})-\theta_2(t,p^2,p^{\prime2})\right],
\end{align}
with $\theta_3(0,p^2,p^{\prime2})=(p^{\prime2}-p^2)d\theta_2(t,p^2,p^{\prime2})/dt|_{t=0}$.
\begin{align}\label{theta4}
&\theta_4(t,p^2,p^{\prime2})=\frac{(p^{\prime2}-p^2)^2}{t}[\theta_2(0,p^2,p^{\prime2})-\theta_2(t,p^2,p^{\prime 2})\nonumber\\
&+(p^2-m_{\pi}^2)\theta_2(0,p^2,m_{\pi}^2)+(p^{\prime2}-m_{\pi}^2)\theta_2(0,m_{\pi}^2,p^{\prime 2})],
\end{align}
$\theta_4$ doesn't appear in Eq. (\ref{pc}).

We have obtained
\begin{align}\label{a93}
&A_{2,0}=\frac{N_cZ_{\pi } }{16\pi^2}\int_0^1 dx\bar{\mathcal{C}}_1(\sigma_1)+\frac{N_cZ_{\pi } }{16\pi^2}\int_0^1 dx\bar{\mathcal{C}}_1(\sigma_2)\nonumber\\
&+\frac{N_cZ_{\pi }}{4\pi^2} \int_0^1 dx \int_0^{1-x}dy  \frac{1}{\sigma_8}\bar{\mathcal{C}}_2(\sigma_8) \nonumber\\
&\times (1-x-y)((p^2+p^{\prime 2})-(x+y)(p^{\prime 2}+p^2-t)),
\end{align}
\begin{align}\label{theta31}
&A_{2,1}=\frac{N_cZ_{\pi }}{4\pi ^2}\int_0^1 dx(\bar{\mathcal{C}}_1(\sigma_1)-\bar{\mathcal{C}}_1(\sigma_2))\nonumber\\
&\times \left(\frac{4 x }{\tau }+\frac{4 x(p^2+p^{\prime 2})}{Q^2 \tau }-\frac{2}{\tau }\right)\nonumber\\
&-\frac{N_c Z_{\pi }}{2\pi^2}\int_0^1 dx \int_0^{1-x} dy(1-x-y)\frac{\bar{\mathcal{C}}_2(\sigma_8)}{\sigma_8}\nonumber\\
&\times (p^{\prime 2}-p^2)\left((x+y)-\frac{(p^{\prime2}+p^2) (1-x-y)}{Q^2}\right) ,
\end{align}
\begin{align}\label{a93}
&A_{2,2}=-\frac{N_cZ_{\pi } }{16\pi^2}\int_0^1 dx\bar{\mathcal{C}}_1(\sigma_1)-\frac{N_cZ_{\pi } }{16\pi^2}\int_0^1 dx\bar{\mathcal{C}}_1(\sigma_2)\nonumber\\
&-\frac{N_cZ_{\pi } }{2\pi^2}\int_0^1 dx x (1-2 x)\bar{\mathcal{C}}_1(\sigma_3)\nonumber\\
&+\frac{N_cZ_{\pi }}{8\pi ^2}\int_0^1 dx(\bar{\mathcal{C}}_1(\sigma_1)-\bar{\mathcal{C}}_1(\sigma_2))\nonumber\\
&\times (p^{\prime 2}-p^2)\left(\frac{ (1-x ) }{Q^2 }-\frac{x (p^{\prime 2}+p^2)}{Q^4 }\right)\nonumber\\
&-\frac{N_cZ_{\pi }}{8\pi ^2}\int_0^1 dx(\bar{\mathcal{C}}_1(\sigma_1)+\bar{\mathcal{C}}_1(\sigma_2))\nonumber\\
&\times  (x -1) \left(\frac{(p^{\prime 2}+p^2)}{Q^2}+1\right)\nonumber\\
&-\frac{N_c Z_{\pi }}{4\pi^2}\int_0^1 dx \int_0^{1-x} dy(\frac{(p^{\prime 2}+p^2)}{Q^2 }+1)\bar{\mathcal{C}}_1(\sigma_8)\nonumber\\
&-\frac{N_c Z_{\pi }}{4\pi^2}\int_0^1 dx \int_0^{1-x} dy(1-x-y)\frac{\bar{\mathcal{C}}_2(\sigma_8)}{\sigma_8}\nonumber\\
&\times (\frac{(p^{'2}+p^2)(1-x-y)}{Q^4}-\frac{  (x+y)}{Q^2})(p^{\prime 2}-p^2)^2,
\end{align}
\begin{align}\label{a1F3}
B_{1,0}&=\frac{ N_c Z_{\pi } }{2\pi ^2}  \int _0^1dx \int _0^{1-x}dy \frac{M m_{\pi}}{\sigma_8}\bar{\mathcal{C}}_2(\sigma_8),
\end{align}
\begin{align}\label{a93}
B_{2,0}&=\frac{N_c Z_{\pi}  }{2\pi ^2} \int_0^1 dx \int_0^{1-x} dy  \nonumber\\
&\times  m_{\pi}M (1-x-y)\frac{1}{\sigma_8}\bar{\mathcal{C}}_2(\sigma_8),
\end{align}
\begin{align}\label{a93}
B_{2,1}&=-\frac{N_cZ_{\pi } }{8\pi^2}\int_0^1 dx\bar{\mathcal{C}}_1(\sigma_1)+\frac{N_cZ_{\pi } }{8\pi^2}\int_0^1 dx\bar{\mathcal{C}}_1(\sigma_2)\nonumber\\
&-\frac{N_cZ_{\pi }}{4\pi^2} \int_0^1 dx \int_0^{1-x}dy  \frac{\bar{\mathcal{C}}_2(\sigma_8) }{\sigma_8}\nonumber\\
&\times m_{\pi}M (1-x-y)\frac{(p^{\prime 2}-p^2)}{Q^2} ,
\end{align}
the FFs \( A_{2,1}(t,p^2,p^{\prime 2}) = -A_{2,1}(t,p^{\prime 2},p^2) \) and \( B_{2,1}(t,p^2,p^{\prime 2}) = -B_{2,1}(t,p^{\prime 2},p^2) \) exhibit antisymmetry. In contrast, the left FFs demonstrate symmetry.

We have confirmed that Eq. (\ref{g}) is numerically equivalent to $-A_{1,1}$ as presented in Eq. (\ref{g1}). However, it should be noted that Eq. (\ref{theta3}) does not exhibit numerical equivalence to $-A_{2,1}$ in Eq. (\ref{theta31}), which differs from the expression given in Eq. (21) of Ref.~\cite{Broniowski:2022iip}, specifically our formulation in Eq. (\ref{theta3}). This discrepancy may arise from the fact that the NJL model does not take into account the gluon distribution in Eq. (\ref{offgpd1}). From Eq. (\ref{theta4}), we can derive the definite form of $\theta_4(t,p^2,p^{\prime 2})$.


%
%

\subsubsection{Impact Parameter Dependent PDFs}
The impact parameter dependent PDFs are defined as
\begin{align}\label{aG9}
q\left(x,\bm{b}_{\perp}^2\right)=\int \frac{d^2\bm{q}_{\perp}}{(2 \pi )^2}e^{-i\bm{b}_{\perp}\cdot \bm{q}_{\perp}} H\left(x,0,-\bm{q}_{\perp}^2\right),
\end{align}
which means the impact parameter dependent PDFs are the Fourier transform of GPDs at $\xi=0$.

\begin{figure}
\centering
\includegraphics[width=0.47\textwidth]{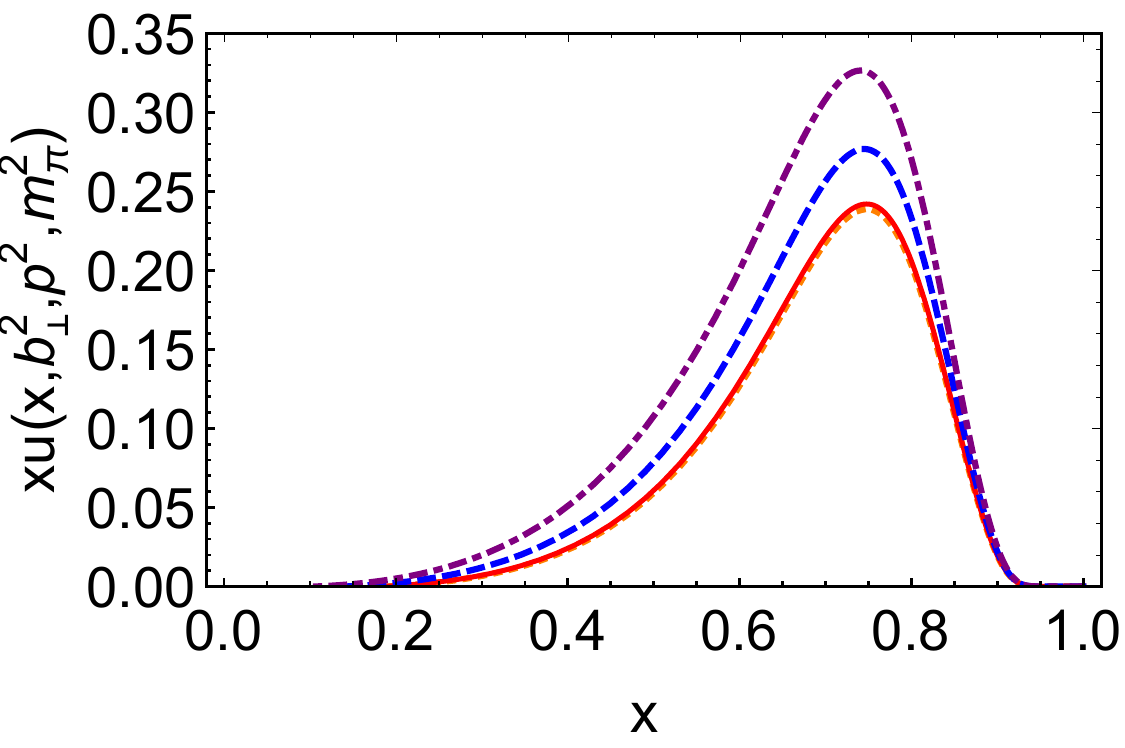}
\qquad
\includegraphics[width=0.47\textwidth]{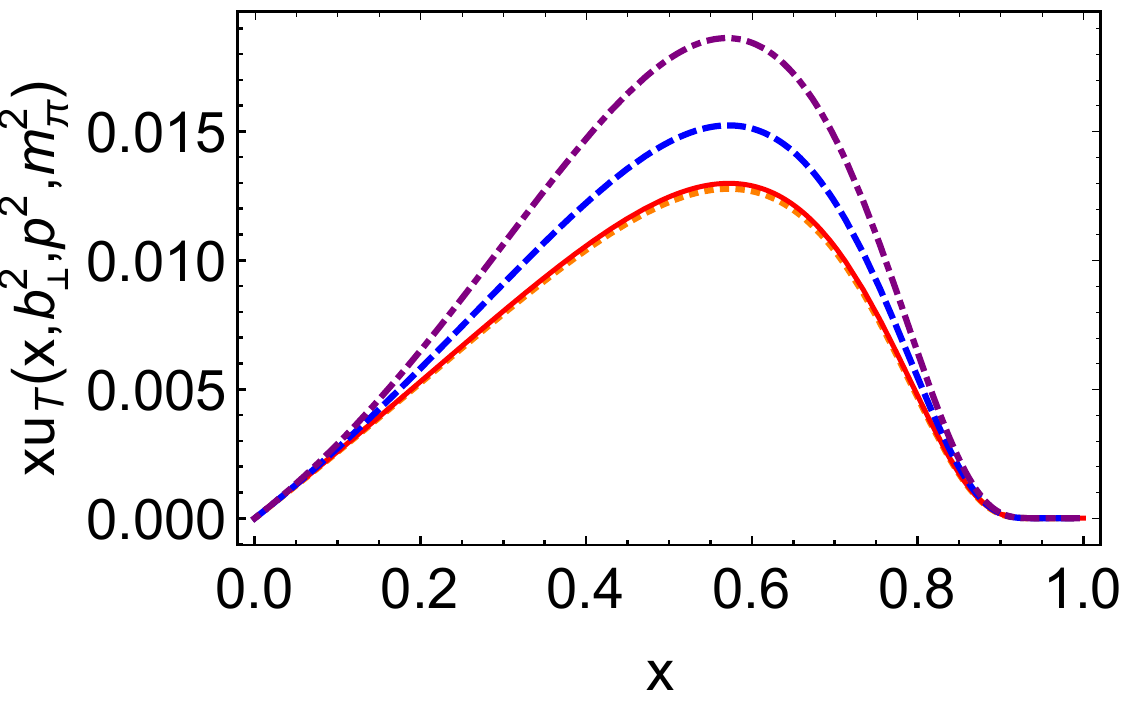}
\caption{Impact parameter space PDFs : upper panel -- $x u\left(x,0.5,p^2,m_{\pi}^2\right)$, the $\delta^2(\bm{b}_{\perp})$ component first line of Eq. (\ref{1ipspdf}) – is suppressed in the image, and lower panel -- $x u_T\left(x,0.5,p^2,m_{\pi}^2\right)$ both panels with $p^2=0$ GeV$^{2}$ --- orange dotted curve, $p^2=0.14^2$ GeV$^{2}$ --- red solid curve, $p^2=0.2$ GeV$^{-2}$ --- blue dashed curve, $p^2=0.4$ GeV$^{2}$ --- purple dot-dashed curve.}\label{qxb2}
\end{figure}

When $\xi=0$ and $t\neq 0$, GPDs become
\begin{align}\label{aG91}
&\quad H(x,0,-\bm{q}_{\perp}^2,p^2,p^{\prime 2})\nonumber\\
&=\frac{N_cZ_{\pi }}{8\pi ^2}  ( \bar{\mathcal{C}}_1(\sigma_1)+\bar{\mathcal{C}}_1(\sigma_2)) \nonumber\\
&+\frac{N_cZ_{\pi }}{4\pi ^2}\int_0^{1-x} d\alpha \left( x (p^2+p^{\prime 2})+ x \bm{q}_{\perp}^2- \bm{q}_{\perp}^2\right)\frac{\bar{\mathcal{C}}_2(\sigma_9)}{\sigma_9},
\end{align}
\begin{align}\label{aG911}
E(x,0,-\bm{q}_{\perp}^2,p^2,p^{\prime 2})=\frac{N_cZ_{\pi }}{2\pi ^2}\int_0^{1-x} d\alpha m_{\pi} M \frac{\bar{\mathcal{C}}_2(\sigma_9)}{\sigma_9},
\end{align}
where $x\in[0, 1]$. After the two-dimensional Fourier transform
\begin{align}\label{1ipspdf}
&\quad u(x,\bm{b}_{\perp}^2,p^2,p^{\prime 2})\nonumber\\
&=\frac{N_cZ_{\pi }}{8\pi ^2} \int \frac{d^2\bm{q}_{\perp}}{(2 \pi )^2}e^{-i\bm{b}_{\perp}\cdot \bm{q}_{\perp}} (\bar{\mathcal{C}}_1(\sigma_1)+\bar{\mathcal{C}}_1(\sigma_2)) \nonumber\\
&+\frac{N_cZ_{\pi }}{32\pi ^3}\int_0^{1-x} d\alpha \int d\tau  \nonumber\\
&\times \frac{(x-1) \left(4-\frac{\bm{b}_{\perp}^2}{\alpha (1-\alpha -x)\tau }\right)+4\alpha   x (1-\alpha-x)\tau (p^2+p^{\prime 2}) }{4 \alpha ^2 \tau ^2 (1-\alpha-x)^2 } \nonumber\\
&\times e^{ -\tau  \left(M^2-(x\alpha p^2+x(1-\alpha-x) p^{\prime 2})\right)} e^{-\frac{\bm{b}_{\perp}^2}{4\tau \left(1-\alpha-x\right) \alpha}},
\end{align}
\begin{align}\label{2ipspdf}
&\quad u_T(x,\bm{b}_{\perp}^2,p^2,p^{\prime 2})\nonumber\\
&=\frac{N_cZ_{\pi }}{16\pi^3}\int_0^{1-x}  d\alpha \int d\tau  \frac{m_{\pi}M }{\alpha\left(1-\alpha-x\right)\tau }\nonumber\\
&\times e^{- \frac{\bm{b}_{\perp}^2}{4\tau (1-\alpha-x)\alpha}}e^{-\tau \left(M^2-(x\alpha p^2+x(1-\alpha-x) p^{\prime 2})\right)},
\end{align}
for \( u\left(x,\bm{b}_{\perp}^2,p^2,p^{\prime 2}\right) \), when integrating over \( \bm{b}_{\perp} \), one can derive the off-shell PDF \( u(x,p^2,p^{\prime 2}) \). The impact parameter space off-shell PDFs are illustrated in Fig. \ref{qxb2}.

%

%
%

\subsubsection{Partons Distribution Functions}
\begin{figure}
\centering
\includegraphics[width=0.47\textwidth]{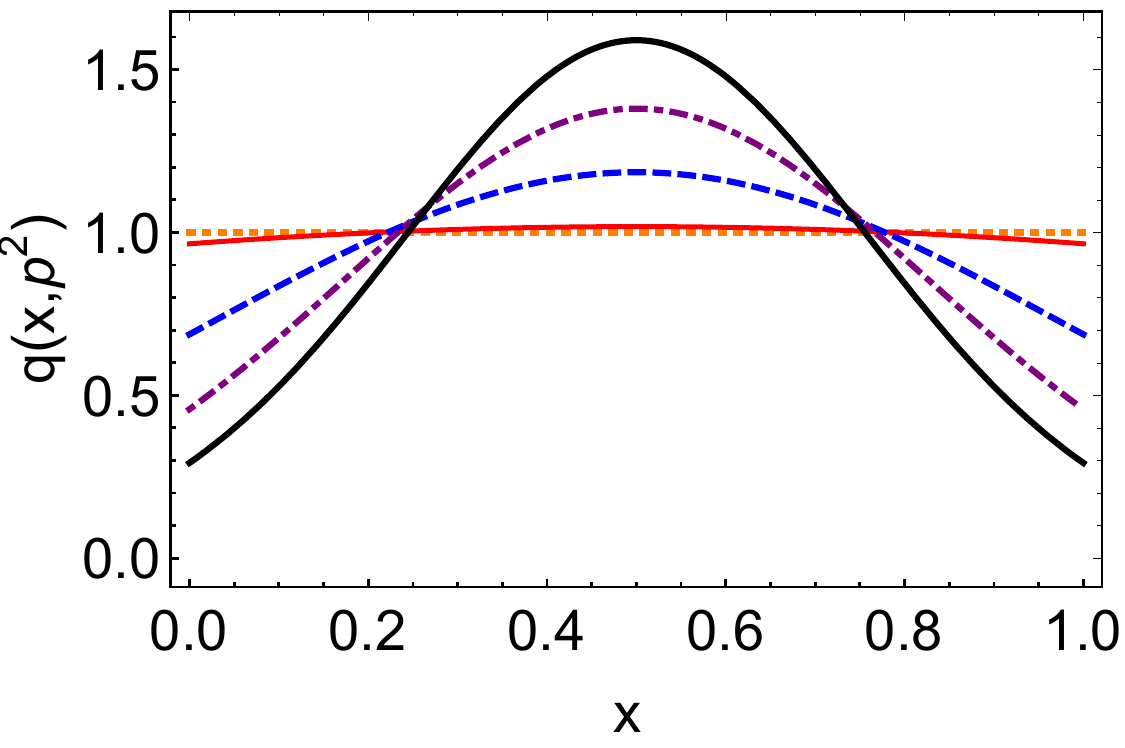}
\caption{The off shell PDFs : upper panel -- $q\left(x,p^2\right)$ with $p^2=0$ GeV$^{2}$ --- orange dotted curve, $p^2=0.14^2$ GeV$^{2}$ --- red solid curve, $p^2=0.2$ GeV$^{-2}$ --- blue dashed curve, $p^2=0.4$ GeV$^{2}$ --- purple dot-dashed curve, $p^2=0.6$ GeV$^{2}$ --- black solid thick curve.}\label{offshellpdf}
\end{figure}

In the forward limit where \( t=0 \) and \( q^2=0 \), under the condition that \( p=p^{\prime} \), we can derive the PDFs
\begin{align}\label{ospdf}
&\quad H(x,0,0,p^2,p^2)=f(x,p^2)\nonumber\\
&=\frac{N_cZ_{\pi }}{4\pi ^2} \bar{\mathcal{C}}_1(\sigma_1)+\frac{N_cZ_{\pi }}{2\pi ^2}  x(1-x) p^2\frac{\bar{\mathcal{C}}_2(\sigma_1)}{\sigma_1},
\end{align}
\begin{align}\label{aG911}
E(x,0,p^2,p^2)=\frac{N_cZ_{\pi }}{2\pi ^2} m_{\pi} M(1-x) \frac{\bar{\mathcal{C}}_2(\sigma_1)}{\sigma_1},
\end{align}
we present the diagrams of the normalized off-shell pion PDFs as illustrated in Fig. \ref{offshellpdf}.
From the diagrams, we observe that in the on-shell case where \( p^2 = m_{\pi}^2 \), the dependence on \( x \) is relatively flat. This observation does not indicate a shortcoming of the NJL model. The NJL model is a Poincaré covariant quantum field theory that shares many low-energy properties with QCD. At high values of \( Q^2 \), it requires evolution; consequently, the dependence on \( x \) will change as a result of this evolution. It is important to note that in the chiral limit, the PDF satisfies \( q(x,0)=1 \), indicating that it remains constant with respect to \( x \).

\section{The pion off-shell TMDs}\label{tmdoffshell}
\begin{figure}
\centering
\includegraphics[width=0.47\textwidth]{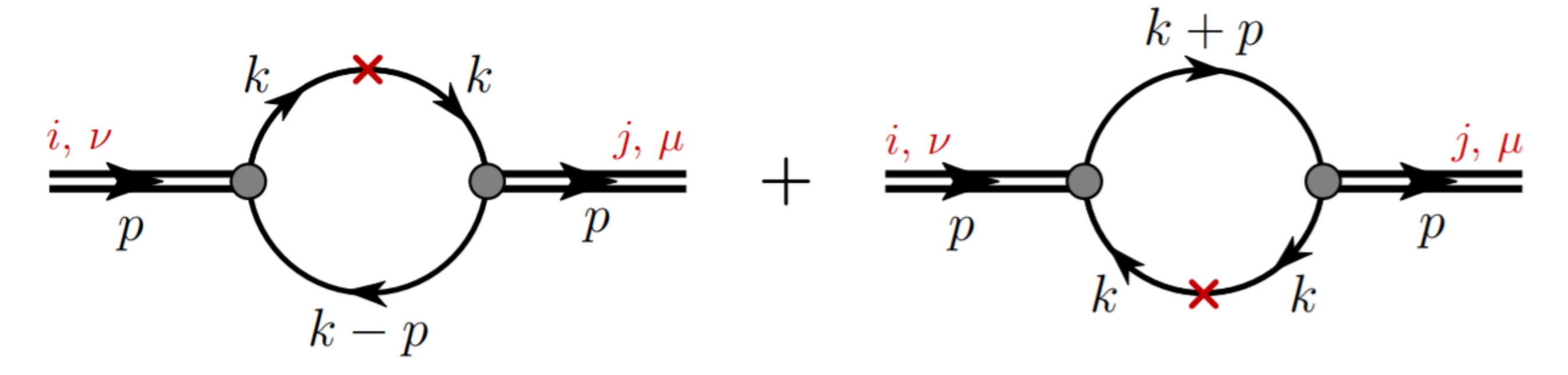}
\caption{Feynman diagrams for the $\pi$ TMDs in the NJL model. The shaded circles represent the $\pi$ Bethe-Salpeter vertex functions and the solid lines the dressed quark propagator. The operator insertion has the form $\gamma^+ \delta(x-\frac{k^+}{p^+})$. The left diagram represent the $u$ quark and the right diagram represent the $d$ quark TMDs in the $\pi$ meson.}\label{TMD}
\end{figure}
\begin{figure}
\centering
\includegraphics[width=0.47\textwidth]{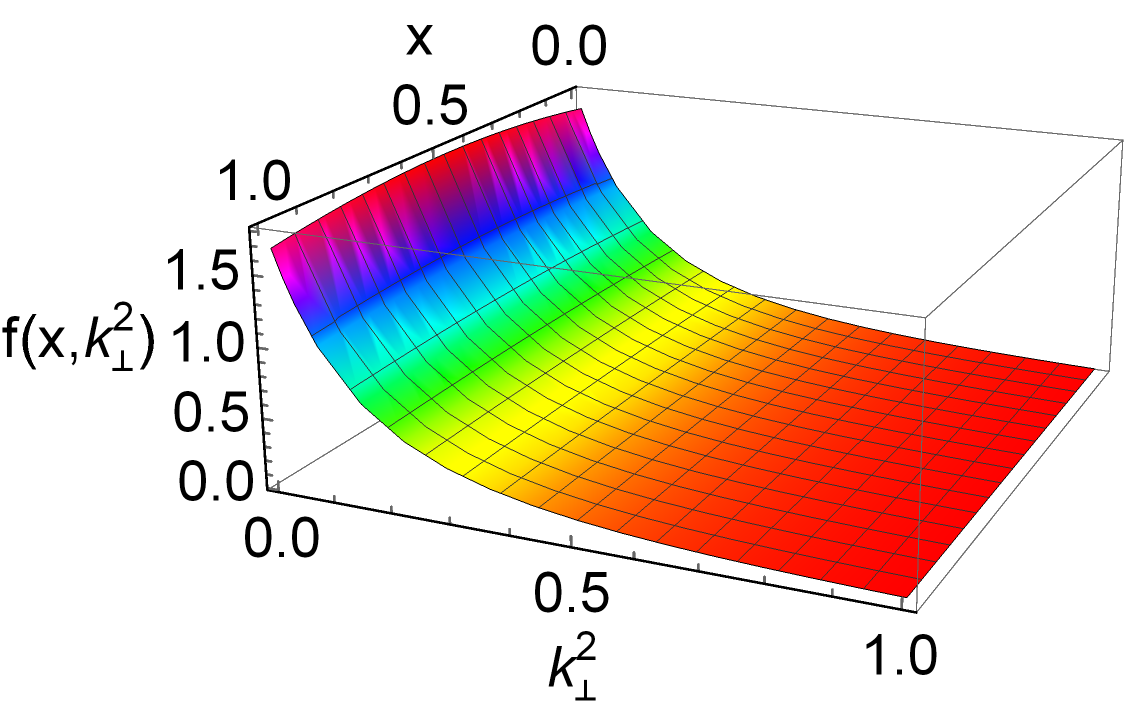}
\qquad
\includegraphics[width=0.47\textwidth]{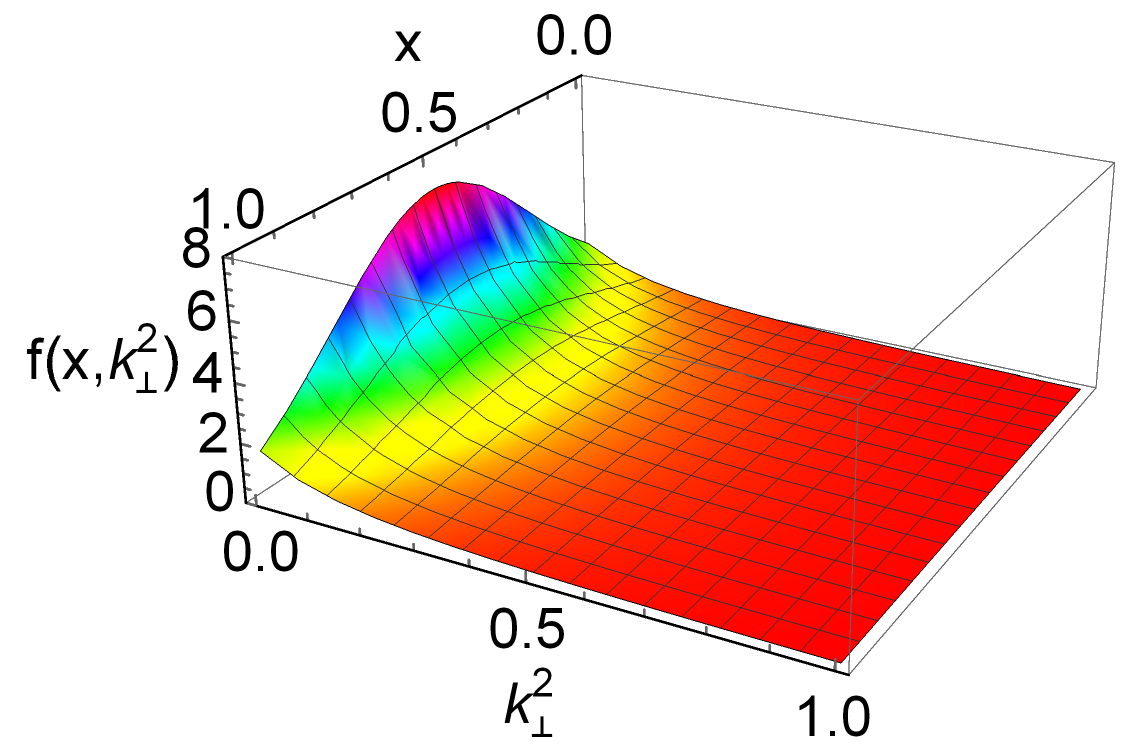}
\caption{Pion TMDs: upper panel -- the on-shell pion TMDs, $f(x,\bm{k}_{\perp}^2,m_{\pi}^2)$, and lower panel -- the off-shell pion TMDs: $f(x,\bm{k}_{\perp}^2,0.4)$.}\label{tmd1}
\end{figure}

The pion TMD is illustrated in Fig. \ref{TMD}. Within the framework of the NJL model, it is defined as follows:
\begin{align}\label{gpddd1}
&\langle\Gamma\rangle(x,\bm{k}_{\perp}^2)=-\frac{i N_c Z_{\pi}}{p^+}\int \frac{\mathrm{d}k^+\mathrm{d}k^-}{(2 \pi )^4}\delta (x-\frac{k^+}{p^+})\nonumber\\
&\times  \text{tr}_{\text{D}}\left[\gamma^5 S \left(k\right)\gamma^+S\left(k\right)\gamma^5 S\left(k-p\right)\right],
\end{align}
where $\text{tr}_{\text{D}}$ denotes a trace over spinor indices. Consequently, we have derived the final expression for the off-shell pion TMD
\begin{align}\label{aG91}
&\quad f(x,\bm{k}_{\perp}^2,p^2)\nonumber\\
&=\frac{N_cZ_{\pi }}{2\pi ^3} \frac{\bar{\mathcal{C}}_2(\sigma_{10})}{\sigma_{10}}+\frac{N_cZ_{\pi }}{4\pi ^3}  x(1-x) p^2\frac{6\bar{\mathcal{C}}_3(\sigma_{10})}{\sigma_{10}^2},
\end{align}
we present a three-dimensional diagram of the on-shell pion TMD and the off-shell pion TMD at \( p^2 = 0.4 \) GeV$^2$ in Fig. \ref{tmd1}. By integrating over \( \bm{k}_{\perp} \), we can derive the off-shell pion PDF as expressed in Eq. (\ref{ospdf}) from the aforementioned equation.
From the diagrams, it is evident that in the upper panel, the unpolarized TMDs exhibit a gradual variation with respect to \(x\). In contrast, the lower panel demonstrates a more pronounced dependence on \(x\) when considering an off-shell pion with \(m_{\pi}^2 = 0.4 \text{ GeV}^2\). The results obtained for the off-shell case align well with findings reported in Refs.~\cite{Frederico:2009fk,Pasquini:2014ppa}, where various quark-based models have been utilized to evaluate unpolarized TMDs.

The dynamics of the NJL model generates the dependence on $\bm{k}_{\perp}$ in our framework. This characteristic is a distinctive feature of the NJL results that is not present in other approaches; the $\bm{k}_{\perp}$ dependence is not introduced through an educated guess. The asymptotic behavior of our results with respect to $\bm{k}_{\perp}$ is 
\begin{align}\label{aG91}
&\quad f(x,\bm{k}_{\perp}^2\rightarrow\infty,p^2)=\frac{3Z_{\pi}}{4\pi^3},
\end{align}
this result is consistent with Ref.~\cite{Noguera:2015iia}, which employed the Pauli-Villars regularization in the NJL model. It also demonstrates that the two different regularization procedures yield qualitatively similar results. This finding aligns with Ref.~\cite{Zhang:2021uak}, which examined pion GPDs in the NJL model using various regularization schemes.

\section{Summary and outlook}\label{excellent}
In this paper, we present our calculations of the off-shell pion generalized parton distributions (GPDs) and transverse momentum dependent parton distribution functions (TMDs) within the framework of the Nambu–Jona-Lasinio (NJL) model. We employ a proper time regularization scheme with an infrared cutoff to effectively simulate confinement.

For the off-shell pion GPDs, when \( p^2 = 0 \) GeV\(^2\), the relative effect is approximately \( 15\% \); at \( p^2 = 0.2 \) GeV\(^2\), the relative effect increases to about \( 25\% \). By analyzing the Mellin moments of GPDs, we can derive the off-shell form factors (FFs) and gravitational form factors. Furthermore, we obtain off-shell pion Parton Distribution Functions (PDFs) from GPDs, with diagrams illustrating that the dependence on \( x \) becomes more pronounced. The results obtained for off-shell conditions are consistent with findings from various quark-based models.


Finally, we investigated the pion off-shell TMDs within the NJL model. Our findings indicate that the TMDs exhibit more pronounced variations with respect to $x$ in the off-shell scenarios, whereas the on-shell TMDs remain relatively flat in this model. Furthermore, the off-shell results demonstrate greater consistency with lattice calculations.

Our findings indicate that the off-shell effects in pion GPDs and TMDs are potentially significant and should not be disregarded. 

In the future, we aim to generalize the off-shell pion GPDs to include kaons. Unlike pions, the isospin symmetry of kaons is broken, allowing us to investigate the differences between the off-shell GPDs of $u$ quarks and $s$ quarks. Furthermore, we will study the off-shell FFs and gravitational FFs of kaons, including a comparison between the off-shell FFs and GFFs of kaons and pions. The off-shell GPDs for pions are associated with the Sullivan process; similarly, considerations regarding the Sullivan process for kaons in an off-shell context at the Electron-Ion Collider are currently underway~\cite{Arrington:2021biu}.



\acknowledgments
Work supported by: the Scientific Research Foundation of Nanjing Institute of Technology (Grant No. YKJ202352).


\appendix
\section{Appendix 1: useful formulae}\label{AppendixT1}

Here we use the gamma-functions ($n\in \mathbb{Z}$, $n\geq 0$)
\begin{subequations}\label{cfun}
\begin{align}
\mathcal{C}_0(z)&:=\int_0^{\infty}\frac{s}{s+z} ds=\int_{\tau_{uv}^2}^{\tau_{ir}^2} d\tau \frac{1}{\tau^2} e^{-\tau z}\,, \\
\mathcal{C}_n(z)&:=(-)^n\frac{\sigma^n}{n!}\frac{d^n}{d\sigma^n}\mathcal{C}_0(\sigma)\,, \\
\bar{\mathcal{C}}_i(z)&:=\frac{1}{z}\mathcal{C}_i(z).
\end{align}
\end{subequations}

The $\sigma$ functions are define as
\begin{subequations}\label{cfun1}
\begin{align}
\sigma_1&=M^2-x(1-x)p^2\,, \\
\sigma_2&=M^2-x(1-x)p^{\prime 2}\,, \\
\sigma_3&=M^2-x(1-x)t=M^2+x(1-x)Q^2\,, \\
\sigma_4&=M^2-\frac{x+\xi}{1+\xi} \frac{1-x}{1+\xi} p^2\,, \\
\sigma_5&=M^2-\frac{x-\xi}{1-\xi}\frac{1-x}{1-\xi} p^{\prime 2}\,, \\
\sigma_6&=M^2-\frac{1}{4}(1+\frac{x}{ \xi })(1-\frac{x}{\xi }) t\,, \\
\sigma_7&=-\alpha \left(\left(\frac{\xi-x}{2\xi}+\alpha\frac{1-\xi}{2\xi}\right) p^2+\left(\frac{\xi+x}{2\xi}-\alpha\frac{1+\xi}{2\xi}\right)p'^2\right)\nonumber\\
&-\left(\frac{\xi+x}{2\xi}-\alpha \frac{1+\xi}{2\xi}\right) (\frac{\xi-x}{2\xi}+\alpha \frac{1-\xi}{2\xi}) t+M^2\,, \\
\sigma_8&=x(x-1)p^{\prime 2}+y(y-1)p^2+xy(p^2+p^{\prime 2})\nonumber\\
&-xyt+M^2\,, \\
\sigma_9&=M^2+(1-\alpha-x) \alpha \bm{q}_{\perp}^2-(x\alpha p^2+x(1-\alpha-x) p^{\prime 2})\,, \\
\sigma_{10}&=\bm{k}_{\perp}^2+M^2+(1-x) x p^2.
\end{align}
\end{subequations}

\bibliographystyle{apsrev4-1}
\bibliography{zhang}

\begin{thebibliography}{65}%
\makeatletter
\providecommand \@ifxundefined [1]{%
 \@ifx{#1\undefined}
}%
\providecommand \@ifnum [1]{%
 \ifnum #1\expandafter \@firstoftwo
 \else \expandafter \@secondoftwo
 \fi
}%
\providecommand \@ifx [1]{%
 \ifx #1\expandafter \@firstoftwo
 \else \expandafter \@secondoftwo
 \fi
}%
\providecommand \natexlab [1]{#1}%
\providecommand \enquote  [1]{``#1''}%
\providecommand \bibnamefont  [1]{#1}%
\providecommand \bibfnamefont [1]{#1}%
\providecommand \citenamefont [1]{#1}%
\providecommand \href@noop [0]{\@secondoftwo}%
\providecommand \href [0]{\begingroup \@sanitize@url \@href}%
\providecommand \@href[1]{\@@startlink{#1}\@@href}%
\providecommand \@@href[1]{\endgroup#1\@@endlink}%
\providecommand \@sanitize@url [0]{\catcode `\\12\catcode `\$12\catcode
  `\&12\catcode `\#12\catcode `\^12\catcode `\_12\catcode `\%12\relax}%
\providecommand \@@startlink[1]{}%
\providecommand \@@endlink[0]{}%
\providecommand \url  [0]{\begingroup\@sanitize@url \@url }%
\providecommand \@url [1]{\endgroup\@href {#1}{\urlprefix }}%
\providecommand \urlprefix  [0]{URL }%
\providecommand \Eprint [0]{\href }%
\providecommand \doibase [0]{http://dx.doi.org/}%
\providecommand \selectlanguage [0]{\@gobble}%
\providecommand \bibinfo  [0]{\@secondoftwo}%
\providecommand \bibfield  [0]{\@secondoftwo}%
\providecommand \translation [1]{[#1]}%
\providecommand \BibitemOpen [0]{}%
\providecommand \bibitemStop [0]{}%
\providecommand \bibitemNoStop [0]{.\EOS\space}%
\providecommand \EOS [0]{\spacefactor3000\relax}%
\providecommand \BibitemShut  [1]{\csname bibitem#1\endcsname}%
\let\auto@bib@innerbib\@empty
\bibitem [{\citenamefont {Zhang}\ \emph
  {et~al.}(2021{\natexlab{a}})\citenamefont {Zhang}, \citenamefont {Cui},
  \citenamefont {Ping},\ and\ \citenamefont {Roberts}}]{Zhang:2020ecj}%
  \BibitemOpen
  \bibfield  {author} {\bibinfo {author} {\bibfnamefont {J.-L.}\ \bibnamefont
  {Zhang}}, \bibinfo {author} {\bibfnamefont {Z.-F.}\ \bibnamefont {Cui}},
  \bibinfo {author} {\bibfnamefont {J.}~\bibnamefont {Ping}}, \ and\ \bibinfo
  {author} {\bibfnamefont {C.~D.}\ \bibnamefont {Roberts}},\ }\href {\doibase
  10.1140/epjc/s10052-020-08791-1} {\bibfield  {journal} {\bibinfo  {journal}
  {Eur. Phys. J. C}\ }\textbf {\bibinfo {volume} {81}},\ \bibinfo {pages} {6}
  (\bibinfo {year} {2021}{\natexlab{a}})},\ \Eprint
  {http://arxiv.org/abs/2009.11384} {arXiv:2009.11384 [hep-ph]} \BibitemShut
  {NoStop}%
\bibitem [{\citenamefont {Puhan}\ \emph {et~al.}(2025)\citenamefont {Puhan},
  \citenamefont {Sharma}, \citenamefont {Kumar},\ and\ \citenamefont
  {Dahiya}}]{Puhan:2025kzz}%
  \BibitemOpen
  \bibfield  {author} {\bibinfo {author} {\bibfnamefont {S.}~\bibnamefont
  {Puhan}}, \bibinfo {author} {\bibfnamefont {S.}~\bibnamefont {Sharma}},
  \bibinfo {author} {\bibfnamefont {N.}~\bibnamefont {Kumar}}, \ and\ \bibinfo
  {author} {\bibfnamefont {H.}~\bibnamefont {Dahiya}},\ }\href@noop {} {\
  (\bibinfo {year} {2025})},\ \Eprint {http://arxiv.org/abs/2504.14982}
  {arXiv:2504.14982 [hep-ph]} \BibitemShut {NoStop}%
\bibitem [{\citenamefont {Zhang}(2024)}]{Zhang:2024adr}%
  \BibitemOpen
  \bibfield  {author} {\bibinfo {author} {\bibfnamefont {J.-L.}\ \bibnamefont
  {Zhang}},\ }\href@noop {} {\  (\bibinfo {year} {2024})},\ \Eprint
  {http://arxiv.org/abs/2409.04105} {arXiv:2409.04105 [hep-ph]} \BibitemShut
  {NoStop}%
\bibitem [{\citenamefont {Muller}\ \emph {et~al.}(1994)\citenamefont {Muller},
  \citenamefont {Robaschik}, \citenamefont {Geyer}, \citenamefont {Dittes},\
  and\ \citenamefont {Horejsi}}]{Mueller:1998fv}%
  \BibitemOpen
  \bibfield  {author} {\bibinfo {author} {\bibfnamefont {D.}~\bibnamefont
  {Muller}}, \bibinfo {author} {\bibfnamefont {D.}~\bibnamefont {Robaschik}},
  \bibinfo {author} {\bibfnamefont {B.}~\bibnamefont {Geyer}}, \bibinfo
  {author} {\bibfnamefont {F.~M.}\ \bibnamefont {Dittes}}, \ and\ \bibinfo
  {author} {\bibfnamefont {J.}~\bibnamefont {Horejsi}},\ }\href {\doibase
  10.1002/prop.2190420202} {\bibfield  {journal} {\bibinfo  {journal} {Fortsch.
  Phys.}\ }\textbf {\bibinfo {volume} {42}},\ \bibinfo {pages} {101} (\bibinfo
  {year} {1994})},\ \Eprint {http://arxiv.org/abs/hep-ph/9812448}
  {arXiv:hep-ph/9812448 [hep-ph]} \BibitemShut {NoStop}%
\bibitem [{\citenamefont {Ji}(1997{\natexlab{a}})}]{Ji:1996nm}%
  \BibitemOpen
  \bibfield  {author} {\bibinfo {author} {\bibfnamefont {X.-D.}\ \bibnamefont
  {Ji}},\ }\href {\doibase 10.1103/PhysRevD.55.7114} {\bibfield  {journal}
  {\bibinfo  {journal} {Phys. Rev. D}\ }\textbf {\bibinfo {volume} {55}},\
  \bibinfo {pages} {7114} (\bibinfo {year} {1997}{\natexlab{a}})},\ \Eprint
  {http://arxiv.org/abs/hep-ph/9609381} {arXiv:hep-ph/9609381} \BibitemShut
  {NoStop}%
\bibitem [{\citenamefont {Radyushkin}(1997)}]{Radyushkin:1997ki}%
  \BibitemOpen
  \bibfield  {author} {\bibinfo {author} {\bibfnamefont {A.~V.}\ \bibnamefont
  {Radyushkin}},\ }\href {\doibase 10.1103/PhysRevD.56.5524} {\bibfield
  {journal} {\bibinfo  {journal} {Phys. Rev.}\ }\textbf {\bibinfo {volume}
  {D56}},\ \bibinfo {pages} {5524} (\bibinfo {year} {1997})},\ \Eprint
  {http://arxiv.org/abs/hep-ph/9704207} {arXiv:hep-ph/9704207 [hep-ph]}
  \BibitemShut {NoStop}%
\bibitem [{\citenamefont {Ji}(1998)}]{Ji:1998pc}%
  \BibitemOpen
  \bibfield  {author} {\bibinfo {author} {\bibfnamefont {X.-D.}\ \bibnamefont
  {Ji}},\ }\href {\doibase 10.1088/0954-3899/24/7/002} {\bibfield  {journal}
  {\bibinfo  {journal} {J. Phys.}\ }\textbf {\bibinfo {volume} {G24}},\
  \bibinfo {pages} {1181} (\bibinfo {year} {1998})},\ \Eprint
  {http://arxiv.org/abs/hep-ph/9807358} {arXiv:hep-ph/9807358 [hep-ph]}
  \BibitemShut {NoStop}%
\bibitem [{\citenamefont {Diehl}(2003)}]{Diehl:2003ny}%
  \BibitemOpen
  \bibfield  {author} {\bibinfo {author} {\bibfnamefont {M.}~\bibnamefont
  {Diehl}},\ }\href {\doibase 10.1016/j.physrep.2003.08.002,
  10.3204/DESY-THESIS-2003-018} {\bibfield  {journal} {\bibinfo  {journal}
  {Phys. Rept.}\ }\textbf {\bibinfo {volume} {388}},\ \bibinfo {pages} {41}
  (\bibinfo {year} {2003})},\ \Eprint {http://arxiv.org/abs/hep-ph/0307382}
  {arXiv:hep-ph/0307382 [hep-ph]} \BibitemShut {NoStop}%
\bibitem [{\citenamefont {Zhang}\ \emph
  {et~al.}(2021{\natexlab{b}})\citenamefont {Zhang}, \citenamefont {Raya},
  \citenamefont {Chang}, \citenamefont {Cui}, \citenamefont {Morgado},
  \citenamefont {Roberts},\ and\ \citenamefont
  {Rodr\'\i{}guez-Quintero}}]{Zhang:2021mtn}%
  \BibitemOpen
  \bibfield  {author} {\bibinfo {author} {\bibfnamefont {J.-L.}\ \bibnamefont
  {Zhang}}, \bibinfo {author} {\bibfnamefont {K.}~\bibnamefont {Raya}},
  \bibinfo {author} {\bibfnamefont {L.}~\bibnamefont {Chang}}, \bibinfo
  {author} {\bibfnamefont {Z.-F.}\ \bibnamefont {Cui}}, \bibinfo {author}
  {\bibfnamefont {J.~M.}\ \bibnamefont {Morgado}}, \bibinfo {author}
  {\bibfnamefont {C.~D.}\ \bibnamefont {Roberts}}, \ and\ \bibinfo {author}
  {\bibfnamefont {J.}~\bibnamefont {Rodr\'\i{}guez-Quintero}},\ }\href
  {\doibase 10.1016/j.physletb.2021.136158} {\bibfield  {journal} {\bibinfo
  {journal} {Phys. Lett. B}\ }\textbf {\bibinfo {volume} {815}},\ \bibinfo
  {pages} {136158} (\bibinfo {year} {2021}{\natexlab{b}})},\ \Eprint
  {http://arxiv.org/abs/2101.12286} {arXiv:2101.12286 [hep-ph]} \BibitemShut
  {NoStop}%
\bibitem [{\citenamefont {Zhang}\ \emph
  {et~al.}(2021{\natexlab{c}})\citenamefont {Zhang}, \citenamefont {Lai},
  \citenamefont {Zong},\ and\ \citenamefont {Ping}}]{Zhang:2021shm}%
  \BibitemOpen
  \bibfield  {author} {\bibinfo {author} {\bibfnamefont {J.-L.}\ \bibnamefont
  {Zhang}}, \bibinfo {author} {\bibfnamefont {M.-Y.}\ \bibnamefont {Lai}},
  \bibinfo {author} {\bibfnamefont {H.-S.}\ \bibnamefont {Zong}}, \ and\
  \bibinfo {author} {\bibfnamefont {J.-L.}\ \bibnamefont {Ping}},\ }\href
  {\doibase 10.1016/j.nuclphysb.2021.115387} {\bibfield  {journal} {\bibinfo
  {journal} {Nucl. Phys. B}\ }\textbf {\bibinfo {volume} {966}},\ \bibinfo
  {pages} {115387} (\bibinfo {year} {2021}{\natexlab{c}})}\BibitemShut
  {NoStop}%
\bibitem [{\citenamefont {Zhang}\ and\ \citenamefont
  {Ping}(2021)}]{Zhang:2021tnr}%
  \BibitemOpen
  \bibfield  {author} {\bibinfo {author} {\bibfnamefont {J.-L.}\ \bibnamefont
  {Zhang}}\ and\ \bibinfo {author} {\bibfnamefont {J.-L.}\ \bibnamefont
  {Ping}},\ }\href {\doibase 10.1140/epjc/s10052-021-09600-z} {\bibfield
  {journal} {\bibinfo  {journal} {Eur. Phys. J. C}\ }\textbf {\bibinfo {volume}
  {81}},\ \bibinfo {pages} {814} (\bibinfo {year} {2021})}\BibitemShut
  {NoStop}%
\bibitem [{\citenamefont {Zhang}\ \emph
  {et~al.}(2022{\natexlab{a}})\citenamefont {Zhang}, \citenamefont {Kang},\
  and\ \citenamefont {Ping}}]{Zhang:2021uak}%
  \BibitemOpen
  \bibfield  {author} {\bibinfo {author} {\bibfnamefont {J.-L.}\ \bibnamefont
  {Zhang}}, \bibinfo {author} {\bibfnamefont {G.-Z.}\ \bibnamefont {Kang}}, \
  and\ \bibinfo {author} {\bibfnamefont {J.-L.}\ \bibnamefont {Ping}},\ }\href
  {\doibase 10.1088/1674-1137/ac57b6} {\bibfield  {journal} {\bibinfo
  {journal} {Chin. Phys. C}\ }\textbf {\bibinfo {volume} {46}},\ \bibinfo
  {pages} {063105} (\bibinfo {year} {2022}{\natexlab{a}})},\ \Eprint
  {http://arxiv.org/abs/2110.06463} {arXiv:2110.06463 [hep-ph]} \BibitemShut
  {NoStop}%
\bibitem [{\citenamefont {Zhang}\ \emph
  {et~al.}(2022{\natexlab{b}})\citenamefont {Zhang}, \citenamefont {Kang},\
  and\ \citenamefont {Ping}}]{Zhang:2022zim}%
  \BibitemOpen
  \bibfield  {author} {\bibinfo {author} {\bibfnamefont {J.-L.}\ \bibnamefont
  {Zhang}}, \bibinfo {author} {\bibfnamefont {G.-Z.}\ \bibnamefont {Kang}}, \
  and\ \bibinfo {author} {\bibfnamefont {J.-L.}\ \bibnamefont {Ping}},\ }\href
  {\doibase 10.1103/PhysRevD.105.094015} {\bibfield  {journal} {\bibinfo
  {journal} {Phys. Rev. D}\ }\textbf {\bibinfo {volume} {105}},\ \bibinfo
  {pages} {094015} (\bibinfo {year} {2022}{\natexlab{b}})},\ \Eprint
  {http://arxiv.org/abs/2204.14032} {arXiv:2204.14032 [hep-ph]} \BibitemShut
  {NoStop}%
\bibitem [{\citenamefont {Deja}\ \emph
  {et~al.}(2023{\natexlab{a}})\citenamefont {Deja}, \citenamefont
  {Martinez-Fernandez}, \citenamefont {Pire}, \citenamefont {Sznajder},\ and\
  \citenamefont {Wagner}}]{Deja:2023tuc}%
  \BibitemOpen
  \bibfield  {author} {\bibinfo {author} {\bibfnamefont {K.}~\bibnamefont
  {Deja}}, \bibinfo {author} {\bibfnamefont {V.}~\bibnamefont
  {Martinez-Fernandez}}, \bibinfo {author} {\bibfnamefont {B.}~\bibnamefont
  {Pire}}, \bibinfo {author} {\bibfnamefont {P.}~\bibnamefont {Sznajder}}, \
  and\ \bibinfo {author} {\bibfnamefont {J.}~\bibnamefont {Wagner}},\ }in\
  \href@noop {} {\emph {\bibinfo {booktitle} {{29th Cracow Epiphany
  Conference}}}}\ (\bibinfo {year} {2023})\ \Eprint
  {http://arxiv.org/abs/2304.03704} {arXiv:2304.03704 [hep-ph]} \BibitemShut
  {NoStop}%
\bibitem [{\citenamefont {Sun}\ and\ \citenamefont {Dong}(2019)}]{Sun:2018ldr}%
  \BibitemOpen
  \bibfield  {author} {\bibinfo {author} {\bibfnamefont {B.-D.}\ \bibnamefont
  {Sun}}\ and\ \bibinfo {author} {\bibfnamefont {Y.-B.}\ \bibnamefont {Dong}},\
  }\href {\doibase 10.1103/PhysRevD.99.016023} {\bibfield  {journal} {\bibinfo
  {journal} {Phys. Rev. D}\ }\textbf {\bibinfo {volume} {99}},\ \bibinfo
  {pages} {016023} (\bibinfo {year} {2019})},\ \Eprint
  {http://arxiv.org/abs/1811.00666} {arXiv:1811.00666 [hep-ph]} \BibitemShut
  {NoStop}%
\bibitem [{\citenamefont {Fu}\ \emph {et~al.}(2022)\citenamefont {Fu},
  \citenamefont {Sun},\ and\ \citenamefont {Dong}}]{Fu:2022bpf}%
  \BibitemOpen
  \bibfield  {author} {\bibinfo {author} {\bibfnamefont {D.}~\bibnamefont
  {Fu}}, \bibinfo {author} {\bibfnamefont {B.-D.}\ \bibnamefont {Sun}}, \ and\
  \bibinfo {author} {\bibfnamefont {Y.}~\bibnamefont {Dong}},\ }\href {\doibase
  10.1103/PhysRevD.106.116012} {\bibfield  {journal} {\bibinfo  {journal}
  {Phys. Rev. D}\ }\textbf {\bibinfo {volume} {106}},\ \bibinfo {pages}
  {116012} (\bibinfo {year} {2022})},\ \Eprint
  {http://arxiv.org/abs/2209.12161} {arXiv:2209.12161 [hep-ph]} \BibitemShut
  {NoStop}%
\bibitem [{\citenamefont {Adhikari}\ \emph {et~al.}(2021)\citenamefont
  {Adhikari}, \citenamefont {Mondal}, \citenamefont {Nair}, \citenamefont {Xu},
  \citenamefont {Jia}, \citenamefont {Zhao},\ and\ \citenamefont
  {Vary}}]{Adhikari:2021jrh}%
  \BibitemOpen
  \bibfield  {author} {\bibinfo {author} {\bibfnamefont {L.}~\bibnamefont
  {Adhikari}}, \bibinfo {author} {\bibfnamefont {C.}~\bibnamefont {Mondal}},
  \bibinfo {author} {\bibfnamefont {S.}~\bibnamefont {Nair}}, \bibinfo {author}
  {\bibfnamefont {S.}~\bibnamefont {Xu}}, \bibinfo {author} {\bibfnamefont
  {S.}~\bibnamefont {Jia}}, \bibinfo {author} {\bibfnamefont {X.}~\bibnamefont
  {Zhao}}, \ and\ \bibinfo {author} {\bibfnamefont {J.~P.}\ \bibnamefont
  {Vary}} (\bibinfo {collaboration} {BLFQ}),\ }\href {\doibase
  10.1103/PhysRevD.104.114019} {\bibfield  {journal} {\bibinfo  {journal}
  {Phys. Rev. D}\ }\textbf {\bibinfo {volume} {104}},\ \bibinfo {pages}
  {114019} (\bibinfo {year} {2021})},\ \Eprint
  {http://arxiv.org/abs/2110.05048} {arXiv:2110.05048 [hep-ph]} \BibitemShut
  {NoStop}%
\bibitem [{\citenamefont {Liu}\ and\ \citenamefont
  {Zahed}(2025)}]{Liu:2025fuf}%
  \BibitemOpen
  \bibfield  {author} {\bibinfo {author} {\bibfnamefont {W.-Y.}\ \bibnamefont
  {Liu}}\ and\ \bibinfo {author} {\bibfnamefont {I.}~\bibnamefont {Zahed}},\
  }\href@noop {} {\  (\bibinfo {year} {2025})},\ \Eprint
  {http://arxiv.org/abs/2503.11959} {arXiv:2503.11959 [hep-ph]} \BibitemShut
  {NoStop}%
\bibitem [{\citenamefont {Zhang}\ and\ \citenamefont
  {Wu}(2025)}]{Zhang:2024plq}%
  \BibitemOpen
  \bibfield  {author} {\bibinfo {author} {\bibfnamefont {J.-L.}\ \bibnamefont
  {Zhang}}\ and\ \bibinfo {author} {\bibfnamefont {J.}~\bibnamefont {Wu}},\
  }\href {\doibase 10.1140/epjc/s10052-024-13682-w} {\bibfield  {journal}
  {\bibinfo  {journal} {Eur. Phys. J. C}\ }\textbf {\bibinfo {volume} {85}},\
  \bibinfo {pages} {13} (\bibinfo {year} {2025})},\ \Eprint
  {http://arxiv.org/abs/2408.13569} {arXiv:2408.13569 [hep-ph]} \BibitemShut
  {NoStop}%
\bibitem [{\citenamefont {Tanisha}\ \emph {et~al.}(2025)\citenamefont
  {Tanisha}, \citenamefont {Puhan}, \citenamefont {Yadav},\ and\ \citenamefont
  {Dahiya}}]{Tanisha:2025qda}%
  \BibitemOpen
  \bibfield  {author} {\bibinfo {author} {\bibnamefont {Tanisha}}, \bibinfo
  {author} {\bibfnamefont {S.}~\bibnamefont {Puhan}}, \bibinfo {author}
  {\bibfnamefont {A.}~\bibnamefont {Yadav}}, \ and\ \bibinfo {author}
  {\bibfnamefont {H.}~\bibnamefont {Dahiya}},\ }\href@noop {} {\  (\bibinfo
  {year} {2025})},\ \Eprint {http://arxiv.org/abs/2505.09213} {arXiv:2505.09213
  [hep-ph]} \BibitemShut {NoStop}%
\bibitem [{\citenamefont {Cheng}\ \emph {et~al.}(2025)\citenamefont {Cheng},
  \citenamefont {Cui}, \citenamefont {Ding}, \citenamefont {Roberts},\ and\
  \citenamefont {Schmidt}}]{Cheng:2024gyv}%
  \BibitemOpen
  \bibfield  {author} {\bibinfo {author} {\bibfnamefont {D.-D.}\ \bibnamefont
  {Cheng}}, \bibinfo {author} {\bibfnamefont {Z.-F.}\ \bibnamefont {Cui}},
  \bibinfo {author} {\bibfnamefont {M.}~\bibnamefont {Ding}}, \bibinfo {author}
  {\bibfnamefont {C.~D.}\ \bibnamefont {Roberts}}, \ and\ \bibinfo {author}
  {\bibfnamefont {S.~M.}\ \bibnamefont {Schmidt}},\ }\href {\doibase
  10.1140/epjc/s10052-025-13782-1} {\bibfield  {journal} {\bibinfo  {journal}
  {Eur. Phys. J. C}\ }\textbf {\bibinfo {volume} {85}},\ \bibinfo {pages} {115}
  (\bibinfo {year} {2025})},\ \Eprint {http://arxiv.org/abs/2409.11568}
  {arXiv:2409.11568 [hep-ph]} \BibitemShut {NoStop}%
\bibitem [{\citenamefont {Wu}\ \emph {et~al.}(2025)\citenamefont {Wu},
  \citenamefont {Cui},\ and\ \citenamefont {Segovia}}]{Wu:2025rto}%
  \BibitemOpen
  \bibfield  {author} {\bibinfo {author} {\bibfnamefont {Q.}~\bibnamefont
  {Wu}}, \bibinfo {author} {\bibfnamefont {Z.-F.}\ \bibnamefont {Cui}}, \ and\
  \bibinfo {author} {\bibfnamefont {J.}~\bibnamefont {Segovia}},\ }\href
  {\doibase 10.1103/9mkx-cn8b} {\bibfield  {journal} {\bibinfo  {journal}
  {Phys. Rev. D}\ }\textbf {\bibinfo {volume} {111}},\ \bibinfo {pages}
  {116023} (\bibinfo {year} {2025})},\ \Eprint
  {http://arxiv.org/abs/2503.07055} {arXiv:2503.07055 [hep-ph]} \BibitemShut
  {NoStop}%
\bibitem [{\citenamefont {Castro}\ \emph {et~al.}(2025)\citenamefont {Castro},
  \citenamefont {Mezrag}, \citenamefont {Morgado~Ch{\'a}vez},\ and\
  \citenamefont {Pire}}]{Castro:2025rpx}%
  \BibitemOpen
  \bibfield  {author} {\bibinfo {author} {\bibfnamefont {A.~R.}\ \bibnamefont
  {Castro}}, \bibinfo {author} {\bibfnamefont {C.}~\bibnamefont {Mezrag}},
  \bibinfo {author} {\bibfnamefont {J.~M.}\ \bibnamefont {Morgado~Ch{\'a}vez}},
  \ and\ \bibinfo {author} {\bibfnamefont {B.}~\bibnamefont {Pire}},\
  }\href@noop {} {\  (\bibinfo {year} {2025})},\ \Eprint
  {http://arxiv.org/abs/2504.02657} {arXiv:2504.02657 [hep-ph]} \BibitemShut
  {NoStop}%
\bibitem [{\citenamefont {Deja}\ \emph
  {et~al.}(2023{\natexlab{b}})\citenamefont {Deja}, \citenamefont
  {Martinez-Fernandez}, \citenamefont {Pire}, \citenamefont {Sznajder},\ and\
  \citenamefont {Wagner}}]{Deja:2023ahc}%
  \BibitemOpen
  \bibfield  {author} {\bibinfo {author} {\bibfnamefont {K.}~\bibnamefont
  {Deja}}, \bibinfo {author} {\bibfnamefont {V.}~\bibnamefont
  {Martinez-Fernandez}}, \bibinfo {author} {\bibfnamefont {B.}~\bibnamefont
  {Pire}}, \bibinfo {author} {\bibfnamefont {P.}~\bibnamefont {Sznajder}}, \
  and\ \bibinfo {author} {\bibfnamefont {J.}~\bibnamefont {Wagner}},\
  }\href@noop {} {\  (\bibinfo {year} {2023}{\natexlab{b}})},\ \Eprint
  {http://arxiv.org/abs/2303.13668} {arXiv:2303.13668 [hep-ph]} \BibitemShut
  {NoStop}%
\bibitem [{\citenamefont {Zhang}(2025)}]{Zhang:2024nxl}%
  \BibitemOpen
  \bibfield  {author} {\bibinfo {author} {\bibfnamefont {J.-L.}\ \bibnamefont
  {Zhang}},\ }\href {\doibase 10.1088/1674-1137/adab61} {\bibfield  {journal}
  {\bibinfo  {journal} {Chin. Phys. C}\ }\textbf {\bibinfo {volume} {49}},\
  \bibinfo {pages} {043104} (\bibinfo {year} {2025})},\ \Eprint
  {http://arxiv.org/abs/2409.19525} {arXiv:2409.19525 [hep-ph]} \BibitemShut
  {NoStop}%
\bibitem [{\citenamefont {Hern{\'a}ndez-Pinto}\ \emph
  {et~al.}(2024)\citenamefont {Hern{\'a}ndez-Pinto}, \citenamefont
  {Guti{\'e}rrez-Guerrero}, \citenamefont {Bedolla},\ and\ \citenamefont
  {Bashir}}]{Hernandez-Pinto:2024kwg}%
  \BibitemOpen
  \bibfield  {author} {\bibinfo {author} {\bibfnamefont {R.~J.}\ \bibnamefont
  {Hern{\'a}ndez-Pinto}}, \bibinfo {author} {\bibfnamefont {L.~X.}\
  \bibnamefont {Guti{\'e}rrez-Guerrero}}, \bibinfo {author} {\bibfnamefont
  {M.~A.}\ \bibnamefont {Bedolla}}, \ and\ \bibinfo {author} {\bibfnamefont
  {A.}~\bibnamefont {Bashir}},\ }\href {\doibase 10.1103/PhysRevD.110.114015}
  {\bibfield  {journal} {\bibinfo  {journal} {Phys. Rev. D}\ }\textbf {\bibinfo
  {volume} {110}},\ \bibinfo {pages} {114015} (\bibinfo {year} {2024})},\
  \Eprint {http://arxiv.org/abs/2410.23813} {arXiv:2410.23813 [hep-ph]}
  \BibitemShut {NoStop}%
\bibitem [{\citenamefont {Bondarenko}\ and\ \citenamefont
  {Slautin}(2025)}]{Bondarenko:2025qch}%
  \BibitemOpen
  \bibfield  {author} {\bibinfo {author} {\bibfnamefont {S.}~\bibnamefont
  {Bondarenko}}\ and\ \bibinfo {author} {\bibfnamefont {M.}~\bibnamefont
  {Slautin}},\ }\href@noop {} {\  (\bibinfo {year} {2025})},\ \Eprint
  {http://arxiv.org/abs/2506.22153} {arXiv:2506.22153 [hep-ph]} \BibitemShut
  {NoStop}%
\bibitem [{\citenamefont {Ji}(1997{\natexlab{b}})}]{Ji:1996ek}%
  \BibitemOpen
  \bibfield  {author} {\bibinfo {author} {\bibfnamefont {X.-D.}\ \bibnamefont
  {Ji}},\ }\href {\doibase 10.1103/PhysRevLett.78.610} {\bibfield  {journal}
  {\bibinfo  {journal} {Phys. Rev. Lett.}\ }\textbf {\bibinfo {volume} {78}},\
  \bibinfo {pages} {610} (\bibinfo {year} {1997}{\natexlab{b}})},\ \Eprint
  {http://arxiv.org/abs/hep-ph/9603249} {arXiv:hep-ph/9603249} \BibitemShut
  {NoStop}%
\bibitem [{\citenamefont {Belitsky}\ and\ \citenamefont
  {Radyushkin}(2005)}]{Belitsky:2005qn}%
  \BibitemOpen
  \bibfield  {author} {\bibinfo {author} {\bibfnamefont {A.~V.}\ \bibnamefont
  {Belitsky}}\ and\ \bibinfo {author} {\bibfnamefont {A.~V.}\ \bibnamefont
  {Radyushkin}},\ }\href {\doibase 10.1016/j.physrep.2005.06.002} {\bibfield
  {journal} {\bibinfo  {journal} {Phys. Rept.}\ }\textbf {\bibinfo {volume}
  {418}},\ \bibinfo {pages} {1} (\bibinfo {year} {2005})},\ \Eprint
  {http://arxiv.org/abs/hep-ph/0504030} {arXiv:hep-ph/0504030} \BibitemShut
  {NoStop}%
\bibitem [{\citenamefont {Mueller}(2014)}]{Mueller:2014hsa}%
  \BibitemOpen
  \bibfield  {author} {\bibinfo {author} {\bibfnamefont {D.}~\bibnamefont
  {Mueller}},\ }\href {\doibase 10.1007/s00601-014-0894-3} {\bibfield
  {journal} {\bibinfo  {journal} {Few Body Syst.}\ }\textbf {\bibinfo {volume}
  {55}},\ \bibinfo {pages} {317} (\bibinfo {year} {2014})},\ \Eprint
  {http://arxiv.org/abs/1405.2817} {arXiv:1405.2817 [hep-ph]} \BibitemShut
  {NoStop}%
\bibitem [{\citenamefont {Sullivan}(1972)}]{Sullivan:1971kd}%
  \BibitemOpen
  \bibfield  {author} {\bibinfo {author} {\bibfnamefont {J.~D.}\ \bibnamefont
  {Sullivan}},\ }\href {\doibase 10.1103/PhysRevD.5.1732} {\bibfield  {journal}
  {\bibinfo  {journal} {Phys. Rev. D}\ }\textbf {\bibinfo {volume} {5}},\
  \bibinfo {pages} {1732} (\bibinfo {year} {1972})}\BibitemShut {NoStop}%
\bibitem [{\citenamefont {Boer}\ and\ \citenamefont
  {Mulders}(1998)}]{Boer:1997nt}%
  \BibitemOpen
  \bibfield  {author} {\bibinfo {author} {\bibfnamefont {D.}~\bibnamefont
  {Boer}}\ and\ \bibinfo {author} {\bibfnamefont {P.~J.}\ \bibnamefont
  {Mulders}},\ }\href {\doibase 10.1103/PhysRevD.57.5780} {\bibfield  {journal}
  {\bibinfo  {journal} {Phys. Rev. D}\ }\textbf {\bibinfo {volume} {57}},\
  \bibinfo {pages} {5780} (\bibinfo {year} {1998})},\ \Eprint
  {http://arxiv.org/abs/hep-ph/9711485} {arXiv:hep-ph/9711485} \BibitemShut
  {NoStop}%
\bibitem [{\citenamefont {Collins}\ \emph {et~al.}(2008)\citenamefont
  {Collins}, \citenamefont {Rogers},\ and\ \citenamefont
  {Stasto}}]{Collins:2007ph}%
  \BibitemOpen
  \bibfield  {author} {\bibinfo {author} {\bibfnamefont {J.~C.}\ \bibnamefont
  {Collins}}, \bibinfo {author} {\bibfnamefont {T.~C.}\ \bibnamefont {Rogers}},
  \ and\ \bibinfo {author} {\bibfnamefont {A.~M.}\ \bibnamefont {Stasto}},\
  }\href {\doibase 10.1103/PhysRevD.77.085009} {\bibfield  {journal} {\bibinfo
  {journal} {Phys. Rev. D}\ }\textbf {\bibinfo {volume} {77}},\ \bibinfo
  {pages} {085009} (\bibinfo {year} {2008})},\ \Eprint
  {http://arxiv.org/abs/0708.2833} {arXiv:0708.2833 [hep-ph]} \BibitemShut
  {NoStop}%
\bibitem [{\citenamefont {Amrath}\ \emph {et~al.}(2005)\citenamefont {Amrath},
  \citenamefont {Bacchetta},\ and\ \citenamefont {Metz}}]{Amrath:2005gv}%
  \BibitemOpen
  \bibfield  {author} {\bibinfo {author} {\bibfnamefont {D.}~\bibnamefont
  {Amrath}}, \bibinfo {author} {\bibfnamefont {A.}~\bibnamefont {Bacchetta}}, \
  and\ \bibinfo {author} {\bibfnamefont {A.}~\bibnamefont {Metz}},\ }\href
  {\doibase 10.1103/PhysRevD.71.114018} {\bibfield  {journal} {\bibinfo
  {journal} {Phys. Rev. D}\ }\textbf {\bibinfo {volume} {71}},\ \bibinfo
  {pages} {114018} (\bibinfo {year} {2005})},\ \Eprint
  {http://arxiv.org/abs/hep-ph/0504124} {arXiv:hep-ph/0504124} \BibitemShut
  {NoStop}%
\bibitem [{\citenamefont {Noguera}\ and\ \citenamefont
  {Scopetta}(2015)}]{Noguera:2015iia}%
  \BibitemOpen
  \bibfield  {author} {\bibinfo {author} {\bibfnamefont {S.}~\bibnamefont
  {Noguera}}\ and\ \bibinfo {author} {\bibfnamefont {S.}~\bibnamefont
  {Scopetta}},\ }\href {\doibase 10.1007/JHEP11(2015)102} {\bibfield  {journal}
  {\bibinfo  {journal} {JHEP}\ }\textbf {\bibinfo {volume} {11}},\ \bibinfo
  {pages} {102} (\bibinfo {year} {2015})},\ \Eprint
  {http://arxiv.org/abs/1508.01061} {arXiv:1508.01061 [hep-ph]} \BibitemShut
  {NoStop}%
\bibitem [{\citenamefont {Lorc\'e}\ \emph {et~al.}(2015)\citenamefont
  {Lorc\'e}, \citenamefont {Pasquini},\ and\ \citenamefont
  {Schweitzer}}]{Lorce:2014hxa}%
  \BibitemOpen
  \bibfield  {author} {\bibinfo {author} {\bibfnamefont {C.}~\bibnamefont
  {Lorc\'e}}, \bibinfo {author} {\bibfnamefont {B.}~\bibnamefont {Pasquini}}, \
  and\ \bibinfo {author} {\bibfnamefont {P.}~\bibnamefont {Schweitzer}},\
  }\href {\doibase 10.1007/JHEP01(2015)103} {\bibfield  {journal} {\bibinfo
  {journal} {JHEP}\ }\textbf {\bibinfo {volume} {01}},\ \bibinfo {pages} {103}
  (\bibinfo {year} {2015})},\ \Eprint {http://arxiv.org/abs/1411.2550}
  {arXiv:1411.2550 [hep-ph]} \BibitemShut {NoStop}%
\bibitem [{\citenamefont {Matevosyan}\ \emph {et~al.}(2012)\citenamefont
  {Matevosyan}, \citenamefont {Bentz}, \citenamefont {Cloet},\ and\
  \citenamefont {Thomas}}]{Matevosyan:2011vj}%
  \BibitemOpen
  \bibfield  {author} {\bibinfo {author} {\bibfnamefont {H.~H.}\ \bibnamefont
  {Matevosyan}}, \bibinfo {author} {\bibfnamefont {W.}~\bibnamefont {Bentz}},
  \bibinfo {author} {\bibfnamefont {I.~C.}\ \bibnamefont {Cloet}}, \ and\
  \bibinfo {author} {\bibfnamefont {A.~W.}\ \bibnamefont {Thomas}},\ }\href
  {\doibase 10.1103/PhysRevD.85.014021} {\bibfield  {journal} {\bibinfo
  {journal} {Phys. Rev. D}\ }\textbf {\bibinfo {volume} {85}},\ \bibinfo
  {pages} {014021} (\bibinfo {year} {2012})},\ \Eprint
  {http://arxiv.org/abs/1111.1740} {arXiv:1111.1740 [hep-ph]} \BibitemShut
  {NoStop}%
\bibitem [{\citenamefont {Puhan}\ \emph {et~al.}(2024)\citenamefont {Puhan},
  \citenamefont {Sharma}, \citenamefont {Kaur}, \citenamefont {Kumar},\ and\
  \citenamefont {Dahiya}}]{Puhan:2023ekt}%
  \BibitemOpen
  \bibfield  {author} {\bibinfo {author} {\bibfnamefont {S.}~\bibnamefont
  {Puhan}}, \bibinfo {author} {\bibfnamefont {S.}~\bibnamefont {Sharma}},
  \bibinfo {author} {\bibfnamefont {N.}~\bibnamefont {Kaur}}, \bibinfo {author}
  {\bibfnamefont {N.}~\bibnamefont {Kumar}}, \ and\ \bibinfo {author}
  {\bibfnamefont {H.}~\bibnamefont {Dahiya}},\ }\href {\doibase
  10.1007/JHEP02(2024)075} {\bibfield  {journal} {\bibinfo  {journal} {JHEP}\
  }\textbf {\bibinfo {volume} {02}},\ \bibinfo {pages} {075} (\bibinfo {year}
  {2024})},\ \Eprint {http://arxiv.org/abs/2310.03464} {arXiv:2310.03464
  [hep-ph]} \BibitemShut {NoStop}%
\bibitem [{\citenamefont {Boglione}\ \emph {et~al.}(2024)\citenamefont
  {Boglione}, \citenamefont {D'Alesio}, \citenamefont {Flore}, \citenamefont
  {Gonzalez-Hernandez}, \citenamefont {Murgia},\ and\ \citenamefont
  {Prokudin}}]{Boglione:2024dal}%
  \BibitemOpen
  \bibfield  {author} {\bibinfo {author} {\bibfnamefont {M.}~\bibnamefont
  {Boglione}}, \bibinfo {author} {\bibfnamefont {U.}~\bibnamefont {D'Alesio}},
  \bibinfo {author} {\bibfnamefont {C.}~\bibnamefont {Flore}}, \bibinfo
  {author} {\bibfnamefont {J.~O.}\ \bibnamefont {Gonzalez-Hernandez}}, \bibinfo
  {author} {\bibfnamefont {F.}~\bibnamefont {Murgia}}, \ and\ \bibinfo {author}
  {\bibfnamefont {A.}~\bibnamefont {Prokudin}},\ }\href {\doibase
  10.1016/j.physletb.2024.138712} {\bibfield  {journal} {\bibinfo  {journal}
  {Phys. Lett. B}\ }\textbf {\bibinfo {volume} {854}},\ \bibinfo {pages}
  {138712} (\bibinfo {year} {2024})},\ \Eprint
  {http://arxiv.org/abs/2402.12322} {arXiv:2402.12322 [hep-ph]} \BibitemShut
  {NoStop}%
\bibitem [{\citenamefont {Aguilar}\ \emph {et~al.}(2019)\citenamefont {Aguilar}
  \emph {et~al.}}]{Aguilar:2019teb}%
  \BibitemOpen
  \bibfield  {author} {\bibinfo {author} {\bibfnamefont {A.~C.}\ \bibnamefont
  {Aguilar}} \emph {et~al.},\ }\href {\doibase 10.1140/epja/i2019-12885-0}
  {\bibfield  {journal} {\bibinfo  {journal} {Eur. Phys. J. A}\ }\textbf
  {\bibinfo {volume} {55}},\ \bibinfo {pages} {190} (\bibinfo {year} {2019})},\
  \Eprint {http://arxiv.org/abs/1907.08218} {arXiv:1907.08218 [nucl-ex]}
  \BibitemShut {NoStop}%
\bibitem [{\citenamefont {Roberts}\ \emph {et~al.}(2021)\citenamefont
  {Roberts}, \citenamefont {Richards}, \citenamefont {Horn},\ and\
  \citenamefont {Chang}}]{Roberts:2021nhw}%
  \BibitemOpen
  \bibfield  {author} {\bibinfo {author} {\bibfnamefont {C.~D.}\ \bibnamefont
  {Roberts}}, \bibinfo {author} {\bibfnamefont {D.~G.}\ \bibnamefont
  {Richards}}, \bibinfo {author} {\bibfnamefont {T.}~\bibnamefont {Horn}}, \
  and\ \bibinfo {author} {\bibfnamefont {L.}~\bibnamefont {Chang}},\ }\href
  {\doibase 10.1016/j.ppnp.2021.103883} {\bibfield  {journal} {\bibinfo
  {journal} {Prog. Part. Nucl. Phys.}\ }\textbf {\bibinfo {volume} {120}},\
  \bibinfo {pages} {103883} (\bibinfo {year} {2021})},\ \Eprint
  {http://arxiv.org/abs/2102.01765} {arXiv:2102.01765 [hep-ph]} \BibitemShut
  {NoStop}%
\bibitem [{\citenamefont {Klevansky}(1992)}]{RevModPhys.64.649}%
  \BibitemOpen
  \bibfield  {author} {\bibinfo {author} {\bibfnamefont {S.~P.}\ \bibnamefont
  {Klevansky}},\ }\href {\doibase 10.1103/RevModPhys.64.649} {\bibfield
  {journal} {\bibinfo  {journal} {Rev. Mod. Phys.}\ }\textbf {\bibinfo {volume}
  {64}},\ \bibinfo {pages} {649} (\bibinfo {year} {1992})}\BibitemShut
  {NoStop}%
\bibitem [{\citenamefont {Buballa}(2005)}]{Buballa:2003qv}%
  \BibitemOpen
  \bibfield  {author} {\bibinfo {author} {\bibfnamefont {M.}~\bibnamefont
  {Buballa}},\ }\href {\doibase 10.1016/j.physrep.2004.11.004} {\bibfield
  {journal} {\bibinfo  {journal} {Phys. Rept.}\ }\textbf {\bibinfo {volume}
  {407}},\ \bibinfo {pages} {205} (\bibinfo {year} {2005})},\ \Eprint
  {http://arxiv.org/abs/hep-ph/0402234} {arXiv:hep-ph/0402234 [hep-ph]}
  \BibitemShut {NoStop}%
\bibitem [{\citenamefont {Zhang}\ \emph {et~al.}(2018)\citenamefont {Zhang},
  \citenamefont {Li},\ and\ \citenamefont {Zong}}]{Zhang:2018ouu}%
  \BibitemOpen
  \bibfield  {author} {\bibinfo {author} {\bibfnamefont {J.-L.}\ \bibnamefont
  {Zhang}}, \bibinfo {author} {\bibfnamefont {C.-M.}\ \bibnamefont {Li}}, \
  and\ \bibinfo {author} {\bibfnamefont {H.-S.}\ \bibnamefont {Zong}},\ }\href
  {\doibase 10.1088/1674-1137/42/12/123105} {\bibfield  {journal} {\bibinfo
  {journal} {Chin. Phys. C}\ }\textbf {\bibinfo {volume} {42}},\ \bibinfo
  {pages} {123105} (\bibinfo {year} {2018})}\BibitemShut {NoStop}%
\bibitem [{\citenamefont {Zhang}\ \emph {et~al.}(2016)\citenamefont {Zhang},
  \citenamefont {Shi}, \citenamefont {Xu},\ and\ \citenamefont
  {Zong}}]{Zhang:2016zto}%
  \BibitemOpen
  \bibfield  {author} {\bibinfo {author} {\bibfnamefont {J.-L.}\ \bibnamefont
  {Zhang}}, \bibinfo {author} {\bibfnamefont {Y.-M.}\ \bibnamefont {Shi}},
  \bibinfo {author} {\bibfnamefont {S.-S.}\ \bibnamefont {Xu}}, \ and\ \bibinfo
  {author} {\bibfnamefont {H.-S.}\ \bibnamefont {Zong}},\ }\href {\doibase
  10.1142/S0217732316500863} {\bibfield  {journal} {\bibinfo  {journal} {Mod.
  Phys. Lett.}\ }\textbf {\bibinfo {volume} {A31}},\ \bibinfo {pages} {1650086}
  (\bibinfo {year} {2016})}\BibitemShut {NoStop}%
\bibitem [{\citenamefont {Cui}\ \emph {et~al.}(2017)\citenamefont {Cui},
  \citenamefont {Zhang},\ and\ \citenamefont {Zong}}]{Cui:2017ilj}%
  \BibitemOpen
  \bibfield  {author} {\bibinfo {author} {\bibfnamefont {Z.-F.}\ \bibnamefont
  {Cui}}, \bibinfo {author} {\bibfnamefont {J.-L.}\ \bibnamefont {Zhang}}, \
  and\ \bibinfo {author} {\bibfnamefont {H.-S.}\ \bibnamefont {Zong}},\ }\href
  {\doibase 10.1038/srep45937} {\bibfield  {journal} {\bibinfo  {journal} {Sci.
  Rep.}\ }\textbf {\bibinfo {volume} {7}},\ \bibinfo {pages} {45937} (\bibinfo
  {year} {2017})}\BibitemShut {NoStop}%
\bibitem [{\citenamefont {Cui}\ \emph {et~al.}(2016)\citenamefont {Cui},
  \citenamefont {Cloët}, \citenamefont {Lu}, \citenamefont {Roberts},
  \citenamefont {Schmidt}, \citenamefont {Xu},\ and\ \citenamefont
  {Zong}}]{Cui:2016zqp}%
  \BibitemOpen
  \bibfield  {author} {\bibinfo {author} {\bibfnamefont {Z.-F.}\ \bibnamefont
  {Cui}}, \bibinfo {author} {\bibfnamefont {I.~C.}\ \bibnamefont {Cloët}},
  \bibinfo {author} {\bibfnamefont {Y.}~\bibnamefont {Lu}}, \bibinfo {author}
  {\bibfnamefont {C.~D.}\ \bibnamefont {Roberts}}, \bibinfo {author}
  {\bibfnamefont {S.~M.}\ \bibnamefont {Schmidt}}, \bibinfo {author}
  {\bibfnamefont {S.-S.}\ \bibnamefont {Xu}}, \ and\ \bibinfo {author}
  {\bibfnamefont {H.-S.}\ \bibnamefont {Zong}},\ }\href {\doibase
  10.1103/PhysRevD.94.071503} {\bibfield  {journal} {\bibinfo  {journal} {Phys.
  Rev. D}\ }\textbf {\bibinfo {volume} {94}},\ \bibinfo {pages} {071503}
  (\bibinfo {year} {2016})},\ \Eprint {http://arxiv.org/abs/1604.08454}
  {arXiv:1604.08454 [nucl-th]} \BibitemShut {NoStop}%
\bibitem [{\citenamefont {Li}\ \emph {et~al.}(2018)\citenamefont {Li},
  \citenamefont {Zhang}, \citenamefont {Yan}, \citenamefont {Huang},\ and\
  \citenamefont {Zong}}]{Li:2018ltg}%
  \BibitemOpen
  \bibfield  {author} {\bibinfo {author} {\bibfnamefont {C.-M.}\ \bibnamefont
  {Li}}, \bibinfo {author} {\bibfnamefont {J.-L.}\ \bibnamefont {Zhang}},
  \bibinfo {author} {\bibfnamefont {Y.}~\bibnamefont {Yan}}, \bibinfo {author}
  {\bibfnamefont {Y.-F.}\ \bibnamefont {Huang}}, \ and\ \bibinfo {author}
  {\bibfnamefont {H.-S.}\ \bibnamefont {Zong}},\ }\href {\doibase
  10.1103/PhysRevD.97.103013} {\bibfield  {journal} {\bibinfo  {journal} {Phys.
  Rev. D}\ }\textbf {\bibinfo {volume} {97}},\ \bibinfo {pages} {103013}
  (\bibinfo {year} {2018})},\ \Eprint {http://arxiv.org/abs/1804.10785}
  {arXiv:1804.10785 [nucl-th]} \BibitemShut {NoStop}%
\bibitem [{\citenamefont {Zhang}\ and\ \citenamefont
  {Wu}(2024)}]{Zhang:2024dhs}%
  \BibitemOpen
  \bibfield  {author} {\bibinfo {author} {\bibfnamefont {J.-L.}\ \bibnamefont
  {Zhang}}\ and\ \bibinfo {author} {\bibfnamefont {J.}~\bibnamefont {Wu}},\
  }\href {\doibase 10.1088/1674-1137/ad2b54} {\bibfield  {journal} {\bibinfo
  {journal} {Chin. Phys. C}\ }\textbf {\bibinfo {volume} {48}},\ \bibinfo
  {pages} {083106} (\bibinfo {year} {2024})},\ \Eprint
  {http://arxiv.org/abs/2402.12757} {arXiv:2402.12757 [hep-ph]} \BibitemShut
  {NoStop}%
\bibitem [{\citenamefont {Chen}\ \emph {et~al.}(2007)\citenamefont {Chen},
  \citenamefont {Detmold},\ and\ \citenamefont {Smigielski}}]{Chen:2006gg}%
  \BibitemOpen
  \bibfield  {author} {\bibinfo {author} {\bibfnamefont {J.-W.}\ \bibnamefont
  {Chen}}, \bibinfo {author} {\bibfnamefont {W.}~\bibnamefont {Detmold}}, \
  and\ \bibinfo {author} {\bibfnamefont {B.}~\bibnamefont {Smigielski}},\
  }\href {\doibase 10.1103/PhysRevD.75.074003} {\bibfield  {journal} {\bibinfo
  {journal} {Phys. Rev. D}\ }\textbf {\bibinfo {volume} {75}},\ \bibinfo
  {pages} {074003} (\bibinfo {year} {2007})},\ \Eprint
  {http://arxiv.org/abs/hep-lat/0612027} {arXiv:hep-lat/0612027} \BibitemShut
  {NoStop}%
\bibitem [{\citenamefont {Broniowski}\ \emph {et~al.}(2008)\citenamefont
  {Broniowski}, \citenamefont {Ruiz~Arriola},\ and\ \citenamefont
  {Golec-Biernat}}]{Broniowski:2007si}%
  \BibitemOpen
  \bibfield  {author} {\bibinfo {author} {\bibfnamefont {W.}~\bibnamefont
  {Broniowski}}, \bibinfo {author} {\bibfnamefont {E.}~\bibnamefont
  {Ruiz~Arriola}}, \ and\ \bibinfo {author} {\bibfnamefont {K.}~\bibnamefont
  {Golec-Biernat}},\ }\href {\doibase 10.1103/PhysRevD.77.034023} {\bibfield
  {journal} {\bibinfo  {journal} {Phys. Rev. D}\ }\textbf {\bibinfo {volume}
  {77}},\ \bibinfo {pages} {034023} (\bibinfo {year} {2008})},\ \Eprint
  {http://arxiv.org/abs/0712.1012} {arXiv:0712.1012 [hep-ph]} \BibitemShut
  {NoStop}%
\bibitem [{\citenamefont {Mezrag}\ \emph {et~al.}(2015)\citenamefont {Mezrag},
  \citenamefont {Chang}, \citenamefont {Moutarde}, \citenamefont {Roberts},
  \citenamefont {Rodr\'\i{}guez-Quintero}, \citenamefont {Sabati\'e},\ and\
  \citenamefont {Schmidt}}]{Mezrag:2014jka}%
  \BibitemOpen
  \bibfield  {author} {\bibinfo {author} {\bibfnamefont {C.}~\bibnamefont
  {Mezrag}}, \bibinfo {author} {\bibfnamefont {L.}~\bibnamefont {Chang}},
  \bibinfo {author} {\bibfnamefont {H.}~\bibnamefont {Moutarde}}, \bibinfo
  {author} {\bibfnamefont {C.~D.}\ \bibnamefont {Roberts}}, \bibinfo {author}
  {\bibfnamefont {J.}~\bibnamefont {Rodr\'\i{}guez-Quintero}}, \bibinfo
  {author} {\bibfnamefont {F.}~\bibnamefont {Sabati\'e}}, \ and\ \bibinfo
  {author} {\bibfnamefont {S.~M.}\ \bibnamefont {Schmidt}},\ }\href {\doibase
  10.1016/j.physletb.2014.12.027} {\bibfield  {journal} {\bibinfo  {journal}
  {Phys. Lett. B}\ }\textbf {\bibinfo {volume} {741}},\ \bibinfo {pages} {190}
  (\bibinfo {year} {2015})},\ \Eprint {http://arxiv.org/abs/1411.6634}
  {arXiv:1411.6634 [nucl-th]} \BibitemShut {NoStop}%
\bibitem [{\citenamefont {Mezrag}\ \emph {et~al.}(2016)\citenamefont {Mezrag},
  \citenamefont {Moutarde},\ and\ \citenamefont
  {Rodriguez-Quintero}}]{Mezrag:2016hnp}%
  \BibitemOpen
  \bibfield  {author} {\bibinfo {author} {\bibfnamefont {C.}~\bibnamefont
  {Mezrag}}, \bibinfo {author} {\bibfnamefont {H.}~\bibnamefont {Moutarde}}, \
  and\ \bibinfo {author} {\bibfnamefont {J.}~\bibnamefont
  {Rodriguez-Quintero}},\ }\href {\doibase 10.1007/s00601-016-1119-8}
  {\bibfield  {journal} {\bibinfo  {journal} {Few Body Syst.}\ }\textbf
  {\bibinfo {volume} {57}},\ \bibinfo {pages} {729} (\bibinfo {year} {2016})},\
  \Eprint {http://arxiv.org/abs/1602.07722} {arXiv:1602.07722 [nucl-th]}
  \BibitemShut {NoStop}%
\bibitem [{\citenamefont {Chavez}\ \emph {et~al.}(2022)\citenamefont {Chavez},
  \citenamefont {Bertone}, \citenamefont {De~Soto~Borrero}, \citenamefont
  {Defurne}, \citenamefont {Mezrag}, \citenamefont {Moutarde}, \citenamefont
  {Rodr\'\i{}guez-Quintero},\ and\ \citenamefont {Segovia}}]{Chavez:2021llq}%
  \BibitemOpen
  \bibfield  {author} {\bibinfo {author} {\bibfnamefont {J.~M.~M.}\
  \bibnamefont {Chavez}}, \bibinfo {author} {\bibfnamefont {V.}~\bibnamefont
  {Bertone}}, \bibinfo {author} {\bibfnamefont {F.}~\bibnamefont
  {De~Soto~Borrero}}, \bibinfo {author} {\bibfnamefont {M.}~\bibnamefont
  {Defurne}}, \bibinfo {author} {\bibfnamefont {C.}~\bibnamefont {Mezrag}},
  \bibinfo {author} {\bibfnamefont {H.}~\bibnamefont {Moutarde}}, \bibinfo
  {author} {\bibfnamefont {J.}~\bibnamefont {Rodr\'\i{}guez-Quintero}}, \ and\
  \bibinfo {author} {\bibfnamefont {J.}~\bibnamefont {Segovia}},\ }\href
  {\doibase 10.1103/PhysRevD.105.094012} {\bibfield  {journal} {\bibinfo
  {journal} {Phys. Rev. D}\ }\textbf {\bibinfo {volume} {105}},\ \bibinfo
  {pages} {094012} (\bibinfo {year} {2022})},\ \Eprint
  {http://arxiv.org/abs/2110.06052} {arXiv:2110.06052 [hep-ph]} \BibitemShut
  {NoStop}%
\bibitem [{\citenamefont {Raya}\ \emph {et~al.}(2022)\citenamefont {Raya},
  \citenamefont {Cui}, \citenamefont {Chang}, \citenamefont {Morgado},
  \citenamefont {Roberts},\ and\ \citenamefont
  {Rodriguez-Quintero}}]{Raya:2021zrz}%
  \BibitemOpen
  \bibfield  {author} {\bibinfo {author} {\bibfnamefont {K.}~\bibnamefont
  {Raya}}, \bibinfo {author} {\bibfnamefont {Z.-F.}\ \bibnamefont {Cui}},
  \bibinfo {author} {\bibfnamefont {L.}~\bibnamefont {Chang}}, \bibinfo
  {author} {\bibfnamefont {J.-M.}\ \bibnamefont {Morgado}}, \bibinfo {author}
  {\bibfnamefont {C.~D.}\ \bibnamefont {Roberts}}, \ and\ \bibinfo {author}
  {\bibfnamefont {J.}~\bibnamefont {Rodriguez-Quintero}},\ }\href {\doibase
  10.1088/1674-1137/ac3071} {\bibfield  {journal} {\bibinfo  {journal} {Chin.
  Phys. C}\ }\textbf {\bibinfo {volume} {46}},\ \bibinfo {pages} {013105}
  (\bibinfo {year} {2022})},\ \Eprint {http://arxiv.org/abs/2109.11686}
  {arXiv:2109.11686 [hep-ph]} \BibitemShut {NoStop}%
\bibitem [{\citenamefont {Mezrag}(2022)}]{Mezrag:2022pqk}%
  \BibitemOpen
  \bibfield  {author} {\bibinfo {author} {\bibfnamefont {C.}~\bibnamefont
  {Mezrag}},\ }\href {\doibase 10.1007/s00601-022-01765-x} {\bibfield
  {journal} {\bibinfo  {journal} {Few Body Syst.}\ }\textbf {\bibinfo {volume}
  {63}},\ \bibinfo {pages} {62} (\bibinfo {year} {2022})},\ \Eprint
  {http://arxiv.org/abs/2207.13584} {arXiv:2207.13584 [hep-ph]} \BibitemShut
  {NoStop}%
\bibitem [{\citenamefont {Broniowski}\ \emph
  {et~al.}(2023{\natexlab{a}})\citenamefont {Broniowski}, \citenamefont
  {Shastry},\ and\ \citenamefont {Ruiz~Arriola}}]{Broniowski:2023his}%
  \BibitemOpen
  \bibfield  {author} {\bibinfo {author} {\bibfnamefont {W.}~\bibnamefont
  {Broniowski}}, \bibinfo {author} {\bibfnamefont {V.}~\bibnamefont {Shastry}},
  \ and\ \bibinfo {author} {\bibfnamefont {E.}~\bibnamefont {Ruiz~Arriola}},\
  }in\ \href@noop {} {\emph {\bibinfo {booktitle} {{29th Cracow Epiphany
  Conference}}}}\ (\bibinfo {year} {2023})\ \Eprint
  {http://arxiv.org/abs/2304.02097} {arXiv:2304.02097 [hep-ph]} \BibitemShut
  {NoStop}%
\bibitem [{\citenamefont {Broniowski}\ \emph
  {et~al.}(2023{\natexlab{b}})\citenamefont {Broniowski}, \citenamefont
  {Shastry},\ and\ \citenamefont {Ruiz~Arriola}}]{Broniowski:2022iip}%
  \BibitemOpen
  \bibfield  {author} {\bibinfo {author} {\bibfnamefont {W.}~\bibnamefont
  {Broniowski}}, \bibinfo {author} {\bibfnamefont {V.}~\bibnamefont {Shastry}},
  \ and\ \bibinfo {author} {\bibfnamefont {E.}~\bibnamefont {Ruiz~Arriola}},\
  }\href {\doibase 10.1016/j.physletb.2023.137872} {\bibfield  {journal}
  {\bibinfo  {journal} {Phys. Lett. B}\ }\textbf {\bibinfo {volume} {840}},\
  \bibinfo {pages} {137872} (\bibinfo {year} {2023}{\natexlab{b}})},\ \Eprint
  {http://arxiv.org/abs/2211.11067} {arXiv:2211.11067 [hep-ph]} \BibitemShut
  {NoStop}%
\bibitem [{\citenamefont {Hellstern}\ \emph {et~al.}(1997)\citenamefont
  {Hellstern}, \citenamefont {Alkofer},\ and\ \citenamefont
  {Reinhardt}}]{Hellstern:1997nv}%
  \BibitemOpen
  \bibfield  {author} {\bibinfo {author} {\bibfnamefont {G.}~\bibnamefont
  {Hellstern}}, \bibinfo {author} {\bibfnamefont {R.}~\bibnamefont {Alkofer}},
  \ and\ \bibinfo {author} {\bibfnamefont {H.}~\bibnamefont {Reinhardt}},\
  }\href {\doibase 10.1016/S0375-9474(97)00412-0} {\bibfield  {journal}
  {\bibinfo  {journal} {Nucl. Phys.}\ }\textbf {\bibinfo {volume} {A625}},\
  \bibinfo {pages} {697} (\bibinfo {year} {1997})},\ \Eprint
  {http://arxiv.org/abs/hep-ph/9706551} {arXiv:hep-ph/9706551 [hep-ph]}
  \BibitemShut {NoStop}%
\bibitem [{\citenamefont {Bentz}\ and\ \citenamefont
  {Thomas}(2001)}]{Bentz:2001vc}%
  \BibitemOpen
  \bibfield  {author} {\bibinfo {author} {\bibfnamefont {W.}~\bibnamefont
  {Bentz}}\ and\ \bibinfo {author} {\bibfnamefont {A.~W.}\ \bibnamefont
  {Thomas}},\ }\href {\doibase 10.1016/S0375-9474(01)01119-8} {\bibfield
  {journal} {\bibinfo  {journal} {Nucl. Phys.}\ }\textbf {\bibinfo {volume}
  {A696}},\ \bibinfo {pages} {138} (\bibinfo {year} {2001})},\ \Eprint
  {http://arxiv.org/abs/nucl-th/0105022} {arXiv:nucl-th/0105022 [nucl-th]}
  \BibitemShut {NoStop}%
\bibitem [{\citenamefont {Choi}\ \emph {et~al.}(2019)\citenamefont {Choi},
  \citenamefont {Frederico}, \citenamefont {Ji},\ and\ \citenamefont
  {de~Melo}}]{Choi:2019nvk}%
  \BibitemOpen
  \bibfield  {author} {\bibinfo {author} {\bibfnamefont {H.-M.}\ \bibnamefont
  {Choi}}, \bibinfo {author} {\bibfnamefont {T.}~\bibnamefont {Frederico}},
  \bibinfo {author} {\bibfnamefont {C.-R.}\ \bibnamefont {Ji}}, \ and\ \bibinfo
  {author} {\bibfnamefont {J.~P. B.~C.}\ \bibnamefont {de~Melo}},\ }\href
  {\doibase 10.1103/PhysRevD.100.116020} {\bibfield  {journal} {\bibinfo
  {journal} {Phys. Rev. D}\ }\textbf {\bibinfo {volume} {100}},\ \bibinfo
  {pages} {116020} (\bibinfo {year} {2019})},\ \Eprint
  {http://arxiv.org/abs/1908.01185} {arXiv:1908.01185 [hep-ph]} \BibitemShut
  {NoStop}%
\bibitem [{\citenamefont {Kohyama}\ \emph {et~al.}(2015)\citenamefont
  {Kohyama}, \citenamefont {Kimura},\ and\ \citenamefont
  {Inagaki}}]{Kohyama:2015hix}%
  \BibitemOpen
  \bibfield  {author} {\bibinfo {author} {\bibfnamefont {H.}~\bibnamefont
  {Kohyama}}, \bibinfo {author} {\bibfnamefont {D.}~\bibnamefont {Kimura}}, \
  and\ \bibinfo {author} {\bibfnamefont {T.}~\bibnamefont {Inagaki}},\ }\href
  {\doibase 10.1016/j.nuclphysb.2015.05.015} {\bibfield  {journal} {\bibinfo
  {journal} {Nucl. Phys. B}\ }\textbf {\bibinfo {volume} {896}},\ \bibinfo
  {pages} {682} (\bibinfo {year} {2015})},\ \Eprint
  {http://arxiv.org/abs/1501.00449} {arXiv:1501.00449 [hep-ph]} \BibitemShut
  {NoStop}%
\bibitem [{\citenamefont {Frederico}\ \emph {et~al.}(2009)\citenamefont
  {Frederico}, \citenamefont {Pace}, \citenamefont {Pasquini},\ and\
  \citenamefont {Salme}}]{Frederico:2009fk}%
  \BibitemOpen
  \bibfield  {author} {\bibinfo {author} {\bibfnamefont {T.}~\bibnamefont
  {Frederico}}, \bibinfo {author} {\bibfnamefont {E.}~\bibnamefont {Pace}},
  \bibinfo {author} {\bibfnamefont {B.}~\bibnamefont {Pasquini}}, \ and\
  \bibinfo {author} {\bibfnamefont {G.}~\bibnamefont {Salme}},\ }\href
  {\doibase 10.1103/PhysRevD.80.054021} {\bibfield  {journal} {\bibinfo
  {journal} {Phys. Rev. D}\ }\textbf {\bibinfo {volume} {80}},\ \bibinfo
  {pages} {054021} (\bibinfo {year} {2009})},\ \Eprint
  {http://arxiv.org/abs/0907.5566} {arXiv:0907.5566 [hep-ph]} \BibitemShut
  {NoStop}%
\bibitem [{\citenamefont {Pasquini}\ and\ \citenamefont
  {Schweitzer}(2014)}]{Pasquini:2014ppa}%
  \BibitemOpen
  \bibfield  {author} {\bibinfo {author} {\bibfnamefont {B.}~\bibnamefont
  {Pasquini}}\ and\ \bibinfo {author} {\bibfnamefont {P.}~\bibnamefont
  {Schweitzer}},\ }\href {\doibase 10.1103/PhysRevD.90.014050} {\bibfield
  {journal} {\bibinfo  {journal} {Phys. Rev. D}\ }\textbf {\bibinfo {volume}
  {90}},\ \bibinfo {pages} {014050} (\bibinfo {year} {2014})},\ \Eprint
  {http://arxiv.org/abs/1406.2056} {arXiv:1406.2056 [hep-ph]} \BibitemShut
  {NoStop}%
\bibitem [{\citenamefont {Arrington}\ \emph {et~al.}(2021)\citenamefont
  {Arrington} \emph {et~al.}}]{Arrington:2021biu}%
  \BibitemOpen
  \bibfield  {author} {\bibinfo {author} {\bibfnamefont {J.}~\bibnamefont
  {Arrington}} \emph {et~al.},\ }\href {\doibase 10.1088/1361-6471/abf5c3}
  {\bibfield  {journal} {\bibinfo  {journal} {J. Phys. G}\ }\textbf {\bibinfo
  {volume} {48}},\ \bibinfo {pages} {075106} (\bibinfo {year} {2021})},\
  \Eprint {http://arxiv.org/abs/2102.11788} {arXiv:2102.11788 [nucl-ex]}
  \BibitemShut {NoStop}%
\end{thebibliography}%


\end{document}